 \documentclass[12pt,preprint]{aastex}

 \usepackage{emulateapj5}

\slugcomment{to appear in the Astrophysical Journal}

\shorttitle{V Sagittae}
\shortauthors{Hachisu and Kato}

\begin{document}

\title{A LIMIT CYCLE MODEL FOR LONG-TERM OPTICAL VARIATIONS OF
V SAGITTAE: \\ THE SECOND EXAMPLE OF ACCRETION WIND EVOLUTION}


\author{Izumi Hachisu}
\affil{Department of Earth Science and Astronomy, 
College of Arts and Sciences, University of Tokyo,
Komaba, Meguro-ku, Tokyo 153-8902, Japan} 
\email{hachisu@chianti.c.u-tokyo.ac.jp}

\and

\author{Mariko Kato}
\affil{Department of Astronomy, Keio University, 
Hiyoshi, Kouhoku-ku, Yokohama 223-8521, Japan} 
\email{mariko@educ.cc.keio.ac.jp}




\begin{abstract}
     V Sagittae shows quasi-periodic 
optical high (soft X-ray off) and low (soft X-ray on) states with 
the total period of $\sim 300$ days.  A binary model is presented 
to explain orbital light curves both for the high and low states
as well as the transition mechanism between them.  The binary model
consists of  a white dwarf (WD), a disk around the WD,
and a lobe-filling main-sequence (MS) companion.
In the optical low state, the mass transfer rate to the WD is small 
and the size of the disk is as small as its Roche lobe size.
In the optical high state, the mass transfer rate to the WD 
exceeds the critical rate of $\sim 1 \times 10^{-6} M_\sun$~yr$^{-1}$,
and the WD blows an optically thick, massive wind.
Surface layers of the disk are blown in the wind and 
the disk surface extends to the companion or over.  
As a result, optical luminosity of the disk
increases by a magnitude because of its large irradiation effect.
The massive wind completely obscures soft X-rays.
This corresponds to the optical high/soft X-ray off state.
The transition between optical high and low states is driven 
by an attenuation of the mass transfer from the secondary.
During the optical high state, the wind from the WD 
hits the companion and strips off its surface layer.
The mass transfer from the companion gradually reduces and stops.
As the mass supply stops, the WD wind weakens and eventually stops.  
The disk shrinks to a Roche lobe size and the optical magnitude drops.
This phase corresponds to the optical low/soft X-ray on state.
Then a rapid mass transfer resumes.
The mass of the WD envelope increases and it blows a wind again.
This cycle is repeated like a limit cycle.
During this intermittent wind phase, the WD can grow in mass 
at the critical rate and eventually reach the Chandrasekhar mass limit.
This process is called ``accretion wind evolution,''
which is a key evolutionary process in a recently developed
evolutionary scenario of Type Ia supernovae.
This evolutionary process was first confirmed in the LMC
supersoft X-ray source RX~J0513.9$-$6951, 
although it commonly occurs in the supersoft X-ray sources 
when the mass transfer rate exceeds the critical rate.
Thus, V~Sge is the second example of accretion wind evolution. 
\end{abstract}




\section{INTRODUCTION}
   V Sagittae is an eclipsing binary system 
exhibiting optical high and low
states \citep{her65}.   V~Sge has been intensively observed
in various bands from radio to X-rays and many groups challenged
to interpret its binary nature, but this peculiar system
still defies explanation
\citep[e.g.,][for recent review and criticism on the binary 
nature]{sma01}.
\par
     There have been many debates on the nature of V~Sge.
\citet{wil86} suggested a cataclysmic variable (CV) nature 
from their observations.  They  
claimed that the observed changes in the line profiles
through the eclipse were similar to the rotational disturbance 
expected from the eclipse of an accretion disk.
However, \citet{woo00} argued that the variability of the emission
lines seen by \citet{wil86} continue all the way around the orbit
and is actually unrelated to the eclipse.  \citet*{rob97} also found
that no evidence of a rotational disturbance in their high-state
spectroscopy of H$\alpha$. 
\par
     \objectname{V Sge} has been suggested to be a supersoft
X-ray source (SSS) \citep{dia95, gre98, pat98, ste98}.
This suggestion stems from similarities 
in the orbital light curve shapes between V~Sge and the LMC supersoft
X-ray source CAL 87 and in the long-term soft X-ray behaviors
between V~Sge and the LMC supersoft X-ray source RX~J0513.9$-$6951.
Soft X-rays were detected only during the optical low states \citep{gre98},
although this is not yet fully established in V~Sge because of their 
sparse period coverage in X-rays.  
\par
     \cite{woo00} argued that the light curves 
of V~Sge can be well fitted by a much simpler model 
\citep[a pair of two stars, see also][]{mad97, loc99, sma01}
and that no one has yet modeled the light
curves or explained the emission-line profiles with the SSS model.
Instead, \citet{loc98dt} and \citet{woo00} proposed a colliding 
wind model of V~Sge both in the optical high and low states
\citep[see, also,][]{gie98, loc98, woo97}.
However, it should be noted that no one successfully proposed
a model of the transition between the optical high and low states.
\par
    The key observational features necessary to be elucidated here 
are: 
(1) V Sge exhibits long-term transitions between 
optical high ($V \sim 11$ and $\sim 180$ days) and low ($V \sim 12$ 
and $\sim 120$ days) states with the total durations of about 
300 days \citep[e.g.,][for the long-term behavior]{sim99}. 
(2) Very soft X-rays are detected only in the long-term optical low
state \citep[e.g.,][]{gre98}.
(3) Recent radio observations indicate a wind mass loss rate of 
$\sim 10^{-5}M_\sun$~yr$^{-1}$ \citep*{loc97, loc99}.  
\par
     These observational characteristics remind us of the LMC supersoft
X-ray source \objectname{RX J0513.9$-$6951} (hereafter RX~J0513).
RX~J0513 shows very similar characteristics: 
(1') quasi-regular transitions between optical
high ($V \sim 16.5$ and $\sim 120$ days) and low ($V \sim 17.3$ 
and $\sim 40$ days) states \citep[e.g.,][]{alc96, cow02},
(2') supersoft X-rays detected only in the optical low states
\citep[e.g.][]{rei96, rei00, sch93, sou96},
and (3') bipolar jets and winds with the velocity of 
$\sim 4000$~km~s$^{-1}$ \citep[e.g.,][]{cra96, hut02, sou96}, 
which is much stronger (faster and denser) in the optical high states.   
Recently, a new transition mechanism between the optical high and
low states of RX~J0513 has been proposed by \citet{hac03ka, hac03kb}. 
Here we try to reproduce the above three
observational features of V~Sge by the same mechanism as in RX~J0513.
\par
     In \S 2, we introduce our binary model for V Sge. 
The orbital light curve modulations in the optical high
and low states are reproduced in \S 3.  We discuss 
various arguments for/against the WD model in \S 4. 
A limit cycle model for the long-term light curve behavior
is formulated in \S 5 and we present our numerical results
in \S 6.  Discussion and conclusions follow in \S7 and \S8,
respectively.

\section{THE BINARY MODEL OF V SAGITTAE}
\begin{figure}
\plotone{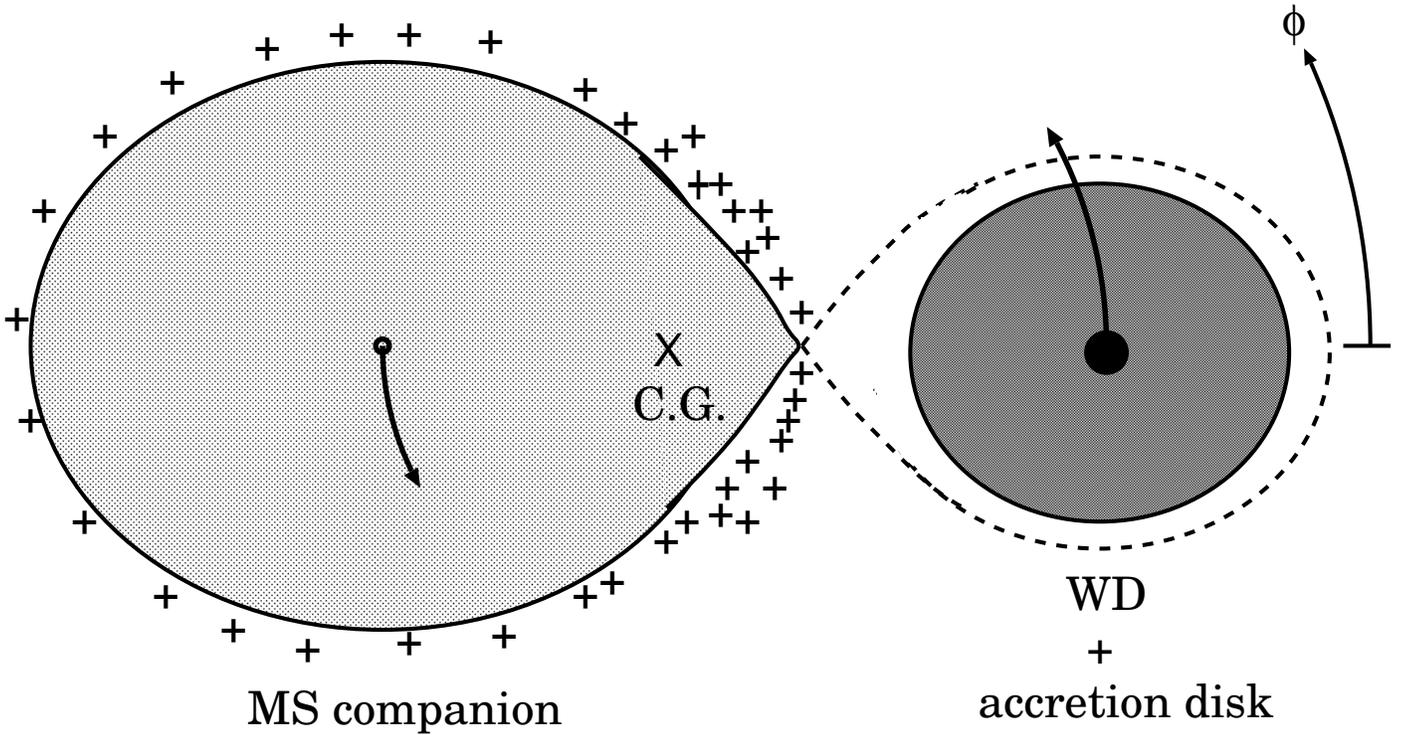}
\caption{
A schematic configuration of our V Sge system in the faint state
is illustrated.
The main-sequence (MS, {\it left}) companion is strongly irradiated 
by the shell-burning white dwarf (WD, {\it right}) so that 
the irradiated hemisphere mainly contributes to the light from
the MS companion.  If the stronger fluorescent \ion{O}{3} line comes
from the surface of the MS companion, it certainly traces closely
the irradiated hemisphere rather than the non-irradiated side.
As a result, the orbital velocity observed is lower than 
that of the MS companion itself.  Crosses around the MS surface 
denote emission of the stronger fluorescent 
\ion{O}{3} line and C.G. indicates the center of gravity of the
binary system.  The azimuthal angle $\phi$ is measured 
from the $x$-axis  of orbital phase 0.5 (the secondary eclipse phase). 
\label{emission_region}}
\end{figure}

     The orbital light curve modulations of V~Sge are 
quite different both in shape and brightness
between the optical high and low states
\citep[e.g.,][]{her65, mad97, sim02}.
In this section, we will show how to construct our binary models
that reproduce orbital modulations in the optical high and low
states.  We adopt a binary system
consisting of a mass-accreting white dwarf (WD), 
a disk around the WD, and a lobe-filling, main-sequence (MS) companion
(see an illustration of Fig. \ref{emission_region}).

\begin{figure}
\plotone{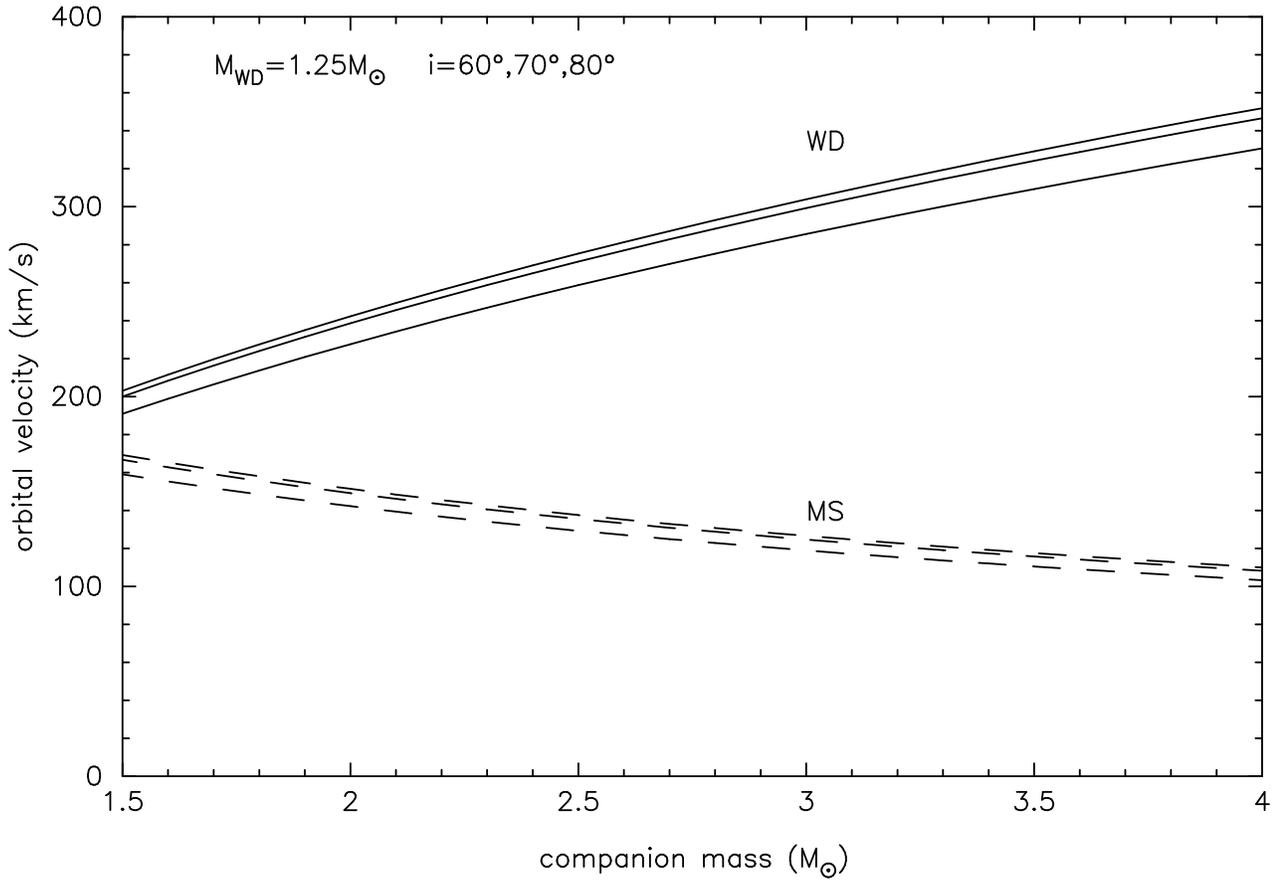}
\caption{
The orbital velocities of each component are plotted against 
the MS companion mass for three inclination angles, i.e.,
$i=80\arcdeg, ~70\arcdeg$, and $60\arcdeg$ (from upper to lower), 
when the WD mass is $M_{\rm WD}= 1.25~M_\sun$.  If we adopt
$K_1 = K_{\rm WD} \sim 300-340$~km~s$^{-1}$, the mass of the MS
companion ranges from $M_2=M_{\rm MS} = 3.0~M_\sun$ to 
$3.5~M_\sun$ and the velocity of the MS companion ranges
from $K_2 = K_{\rm MS} = 130$~km~s$^{-1}$ to 110~km~s$^{-1}$.
\label{orbital_velocity}}
\end{figure}

\placefigure{emission_region}
\placefigure{orbital_velocity}

\subsection{Binary parameters}
     In this section, we assume the WD mass of $M_{\rm WD} 
\approx 1.25~M_\sun$ taking the results of \S 6 in advance, i.e.,
\begin{equation}
M_1 = M_{\rm WD}= 1.25~M_\sun.
\label{WD_mass}
\end{equation}
We adopt the ephemeris by \citet{pat98}, i.e.,
\begin{eqnarray}
t(\mbox{HJD}) &=& 2,437,889.916 + 0.5141967 \times E \cr
 & & - 9.2 \times 10^{-11} E^2,
\label{new_ephemeris}
\end{eqnarray}
at eclipse minima \citep[see also,][]{loc99, mad97, sim96b, sma95}.
\par
     The secondary mass is determined as follows:
\citet{her65} estimated each component mass from
the observed radial velocities.  They assumed that 
the weaker \ion{O}{3} fluorescent line correctly traces 
the motion of the primary while the stronger \ion{O}{3} 
fluorescent line follows the motion of the secondary, i.e., 
$M_1 = (0.74 \pm 0.19) ~M_\sun / \sin^3 i$ 
from $K_1 \sim (320 \pm 28)$~km~s$^{-1}$ and
$M_2 = (2.80 \pm 0.65) ~M_\sun / \sin^3 i$
from $K_2 \sim (85 \pm 9)$~km~s$^{-1}$ \citep{sma01},
where $i$ is the inclination angle of the binary.
Radial velocities of these two sharp lines are oscillating in the
same period of the eclipsing binary and in anti-phase each other.
Moreover, gamma velocities of these two lines are almost coincident.
We do not think that the other lines such as H and \ion{He}{2} are
correct tracers of binary components because they are broad and 
their gamma velocities are not consistent with those 
of \ion{O}{3} fluorescent lines.
\par
     Here, we reexamine the companion mass.  
We assume, following Herbig et al.'s paper, 
that the weaker \ion{O}{3} fluorescent
line correctly traces the motion of the primary,
i.e., $K_1 \sim 300-340$~km~s$^{-1}$ \citep{her65}, but do not
think that the stronger \ion{O}{3} fluorescent line correctly follows
the motion of the secondary.  The reason is as follows:
When the companion star is strongly irradiated by a luminous
(hydrogen shell-burning) WD, 
its surface temperature is much higher at the
irradiated hemisphere than that of the other side
(see Fig. \ref{emission_region}).  Probably the stronger 
\ion{O}{3} fluorescent line should trace more closely the irradiated
hemisphere of the cool component.  As a result, it is very likely
that the observed orbital velocity of the stronger \ion{O}{3}
fluorescent line 
is a bit slower than the motion of the cool component
as seen in Figure \ref{emission_region}.
If this is the case, the observed velocity of 
$K_2 = 85 \pm 9$~km~s$^{-1}$ does not represent the true radial 
velocity of the cool companion.
\par
     Figure \ref{orbital_velocity} shows the 
relation between the orbital radial velocities and 
the MS companion mass derived from Kepler motion with the period
of 0.514 days.  This figure shows the secondary mass between 
$M_{\rm MS}= 3.0$ and $3.5~M_\sun$ 
for $K_1 \sim 300-340$~km~s$^{-1}$, and 
the orbital radial velocity of $K_2 = 130-110$~km~s$^{-1}$. 
Here, we adopt the mass of $M_2= M_{\rm MS}= 3.0 ~M_\sun$. 
Then the separation is $a= 4.375~R_\sun$, the radii of the
effective Roche lobes are $R_1^* = 1.34 ~R_\sun$ and
$R_2^* = R_2 = 2.0 ~R_\sun$.  
The radius of $3.0 ~M_\sun$ zero-age main-sequence
(ZAMS) star is $R_{\rm ZAMS}= 2.5 ~R_\sun$ \citep[e.g.,][]{bre93},
 so that the secondary
is a main-sequence (MS) star but it is undersized and probably 
underluminous because of thermal imbalance 
by the Kelvin-Helmholtz timescale mass-transfer.  We assume
$T_{\rm MS, org}= 12,000$~K for non-irradiated temperature of the MS
companion.  The inclination angle is determined by light curve 
fitting as described below.
It should be noted here that the thermal timescale mass-transfer
is realized for the mass ratio $M_2/M_1 \gtrsim 1$ and 
the mass transfer rate can reach $\sim 10^{-5} M_\sun$~yr$^{-1}$ or
more \citep[e.g.][]{lih97}.
\par
     Our binary models for numerical calculation are illustrated 
in Figure \ref{w125m30configure80}.
A circular orbit is assumed.  We also assume that the photospheric
surfaces of the WD, the MS companion, and the disk  
emit photons as a blackbody at a local temperature 
that varies with position.  The numerical method adopted here
was fully described in \citet{hac01kb}.  
%
%

\placefigure{w125m30configure80}

\subsection{Flaring-up disk and its irradiation effect}
     The irradiation effects both of the disk and of the MS companion
play an essential role in the light curve modeling. 
For the basic structure of the accretion disk,
we assume an axisymmetric structure with the size and thickness of
\begin{equation}
R_{\rm disk} = \alpha R_1^*,
\label{accretion-disk-size}
\end{equation}
and
\begin{equation}
h = \beta R_{\rm disk} \left({{\varpi} 
\over {R_{\rm disk}}} \right)^\nu,
\label{flaring-up-disk}
\end{equation}
where $R_{\rm disk}$ is the outer edge of the accretion disk,
$R_1^*$ is the effective radius of the inner critical Roche lobe 
for the WD component,
$h$ is the height of the surface from the equatorial plane, and
$\varpi$ is the distance from the symmetric axis.
\par
     Since the orbital light curve modulations of V~Sge
show a nonaxisymmetric shape, we introduce an asymmetric 
configuration of the accretion disk as done by \citet*{sch97}.
They explained an asymmetric feature in the orbital light curves 
of the LMC luminous supersoft X-ray source CAL 87.  The cause is that
the gas stream from the companion hits the edge of the accretion disk
and makes a vertical spray when the mass
transfer rate is as high as in the luminous supersoft X-ray sources
($\sim 1 \times 10^{-7} M_\sun$~yr$^{-1}$).
We have also introduced the same type of asymmetry
of the accretion disk as Schandl et al.'s, i.e.,
\begin{equation}
\zeta_{\rm edge} = \left\{ \begin{array}{ll}
 1, &\mbox{~for~} \phi_4 \le \phi < 2 \pi \cr
 \zeta_{\rm low}+(1-\zeta_{\rm low}){{\phi-\phi_3} \over {\phi_4-\phi_3}} 
&\mbox{~for~} \phi_3 \le \phi < \phi_4 \cr
 \zeta_{\rm low}, &\mbox{~for~} \phi_2 \le \phi < \phi_3 \cr
 \zeta_{\rm low}+(1-\zeta_{\rm low}){{\phi-\phi_2} \over {\phi_1-\phi_2}} 
&\mbox{~for~} \phi_1 \le \phi < \phi_2 \cr
 1, &\mbox{~for~} 0 \le \phi < \phi_1 
\end{array}
\right.
\end{equation}
and
\begin{equation}
{{z} \over {h}} = \left\{ \begin{array}{ll}
\zeta_{\rm low}+(\zeta_{\rm edge}-\zeta_{\rm low})
{{\xi-f_a} \over {1-f_a}},
& \mbox{~for~} \xi \ge f_a, \cr
\zeta_{\rm low}
& \mbox{~for~} \xi < f_a, \cr
\end{array}
\right.
\end{equation}
where 
\begin{equation}
\xi= {{\varpi} \over {R_{\rm disk}}},
\end{equation}
and $\zeta_{\rm low}$ is a parameter specifying the degree of asymmetry,
$\phi_3$ is the starting angle of the vertical spray, 
at $\phi_4$ the vertical spray reaches its maximum height, 
this maximum height continues until the angle of $\phi_1$,
the elevated rim disappear at the angle of $\phi_2$,
$f_a$ is the starting radius ratio 
from where the vertical spray is prominent, 
$z$ is the height of the disk surface from the equatorial plane
for the asymmetric case.  The azimuthal angle is measured from
the phase at the WD component in front of the MS companion,
i.e., from the phase at the secondary eclipse 
(see Fig. \ref{emission_region}).
Here, we assume that the accretion disk is symmetric inside 
$\varpi < 0.8 R_{\rm disk}$ ($f_a=0.8$).
The other parameters of $\zeta_{\rm low}$, $\phi_1$, $\phi_2$,
$\phi_3$, and $\phi_4$ are determined from the light curve 
fitting.  Such an example of the accretion disk is shown 
in Figure \ref{w125m30configure80}.  Here, we adopt
$\alpha=6.0$, $\beta=0.24$, $\nu=1$,
$\zeta_{\rm low}=0.1$, $\phi_1=(21/16)\pi$, $\phi_2=(25/16)\pi$,
$\phi_3=(5/16)\pi$, and $\phi_4=(10/16)\pi$ for configuration (a) but
$\alpha=1.15$, $\beta=0.4$, $\nu=2$,
$\zeta_{\rm low}=0.25$, $\phi_1=(23/16)\pi$, $\phi_2=(27/16)\pi$,
$\phi_3=(3/16)\pi$, and $\phi_4=(7/16)\pi$ for configuration (b).
The irradiation efficiency is assumed to be $\eta_{\rm DK}= 0.5$,
i.e., 50\% of the irradiated energy is emitted but the residual
50\% is converted into thermal energy.
\par
     Recent analysis on the recurrent nova CI Aql 2000 outburst 
\citep{hac03ka} and the LMC supersoft X-ray source RX~J0513
\citep{hac03kb} indicates
that the irradiation area of the disk is blown in the wind
and extends up to the companion star or over ($\alpha \sim 3$ 
or over) as schematically shown in Figure \ref{opaque_disk}.
This physically means that (1) the surface layer of 
the accretion disk is blown off and accelerated up to, at least,
several hundred km~s$^{-1}$ like a free stream going outward,
and that (2) the optically thick part of the disk surface ends 
somewhere far outside the Roche lobe as seen in Figures 
\ref{w125m30configure80}a and \ref{opaque_disk}a.
   Here, we adopt $\nu=1$ during the massive wind
phase but  $\nu=2$ in the no massive wind phase
\citep{hac03ka, hac03kb}. 
This $\varpi$-square law mimics the effect of 
flaring-up at the rim of the disk by vertical spray as discussed
by \citet{sch97} for supersoft X-ray sources.
\par
     The accretion luminosity of the disk is also numerically
included, although its contribution to the $V$-light is much 
smaller compared with that of the irradiation effect 
\citep*[see discussion of][]{hac01ka, hac01kb, hac03a}.
The temperature of the outer rim of the disk
is assumed here to be 10000~K.
We have checked that the light curves are hardly changed
even if 8000~K or 12000~K is adopted as an outer rim
temperature instead of 10000~K.

\placefigure{opaque_disk}

\subsection{Irradiated MS companion}
     The MS companion is also irradiated by the WD.
As already shown by \citet{sma01}, the irradiation
effect alone cannot closely reproduce the orbital modulations.
\citet{sch97} adopted a redistribution of the irradiated energy
with a diffusion angle of $45\arcdeg$ to reproduce a
relatively round shape of orbital modulations just before/after
the eclipse minima.   Here, we also adopt the same type of
redistributions of irradiated energy, but with a bit larger
diffusion angle of $60\arcdeg$ instead of $45\arcdeg$.
The irradiation efficiency is the same as that of \citet{sch97},
i.e., assumed to be $\eta_{\rm MS}= 0.5$, i.e., 50\% of the irradiated
energy is emitted but the residual 50\% is converted into
thermal energy.  The distribution of surface temperature is
determined by the constraint that the total of the emitted
energy is equal to the sum of 50\% of the irradiated energy
and the emitted energy with the original surface temperature
of 12,000~K.

\placefigure{vmag_w125m30_orbital}

\section{ORBITAL LIGHT CURVES IN THE OPTICAL HIGH AND LOW STATES}
\begin{figure}
\plotone{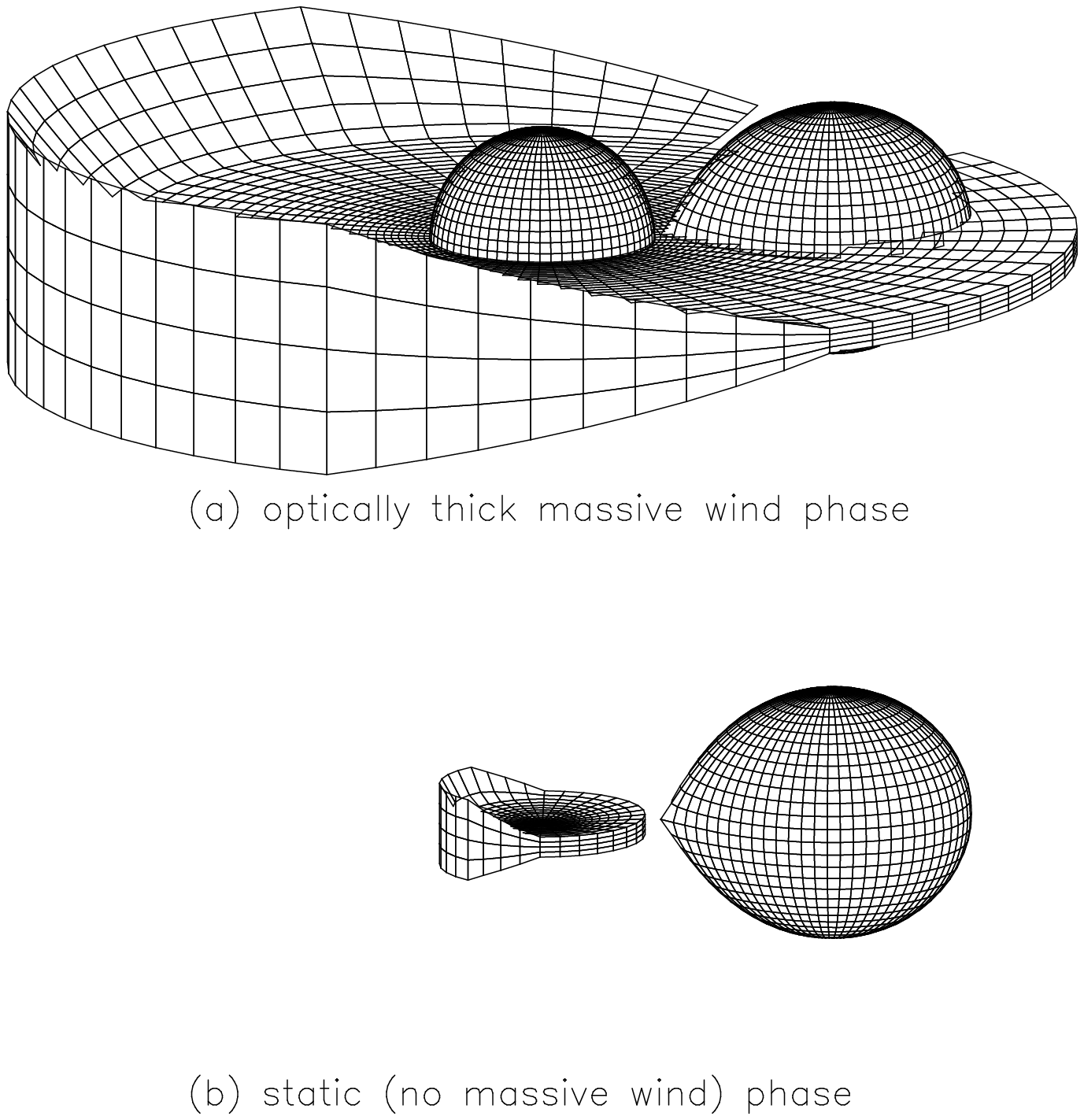}
\caption{
Configurations of our V Sge model are illustrated: (a)
in the massive wind phase (optical high/soft X-ray off), 
and (b) in the static shell-burning phase, i.e.,
the rapid mass accretion phase 
after the wind stops (optical low/soft X-ray on).
The cool component ({\it right}) is a main-sequence
(MS) companion ($3.0 M_\odot$) filling up its inner critical 
Roche lobe.  
The north and south polar areas of the cool component are 
irradiated by the hot component 
($1.25~M_\odot$ white dwarf, {\it left}).
The separation is $a= 4.375 R_\odot$; 
the effective radii of the inner critical Roche lobes are
$R_1^*= 1.34 R_\odot$, and $R_2^*= R_2= 2.0 R_\odot$, 
for the primary WD and the secondary MS companion, respectively.
The surface layer of the disk is blown in the wind, like a free
stream.  This disk surface stream is optically thick near 
the original disk region but becomes optically thin far outside
the original disk due to geometrical dilution effect.  We regard
the transition place from optically thick to thin as the outer edge
of the disk in the massive wind phase (see also Fig. \ref{opaque_disk}a).
This edge extends over the MS companion as shown in the figure.
The disk shrinks to a normal size (here $1.15$ times the Roche lobe 
size) in several orbital periods after the wind stops.  A rapid mass
accretion resumes to make a spray around the disk edge in more several
orbital periods.  The inclination angle is $i= 77\arcdeg$ 
and the azimuthal angle is $\phi = 80\arcdeg$.  Here, the 
azimuthal angle is measured from the angle at the 
secondary eclipse, i.e., at the phase with the WD component
in front of the MS companion.
\label{w125m30configure80}}
\end{figure}

\begin{figure}
\plotone{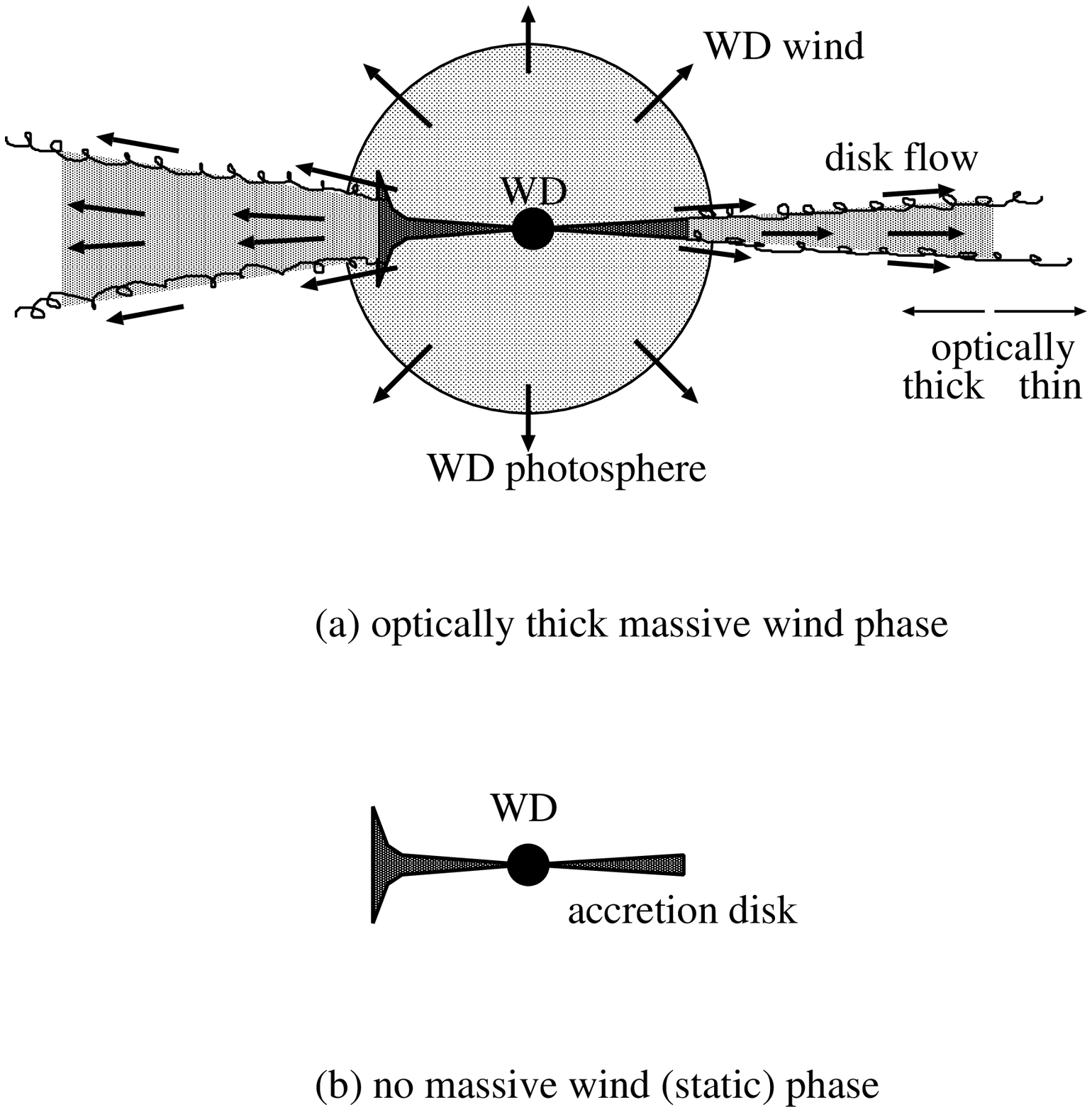}
\caption{
Optically thick parts of the disk are illustrated: (a)
in the massive wind phase (optical high/soft X-ray off), 
and (b) in the static shell-burning phase 
(optical low/soft X-ray on).  See text for more details.
\label{opaque_disk}}
\end{figure}

\begin{figure}
\plotone{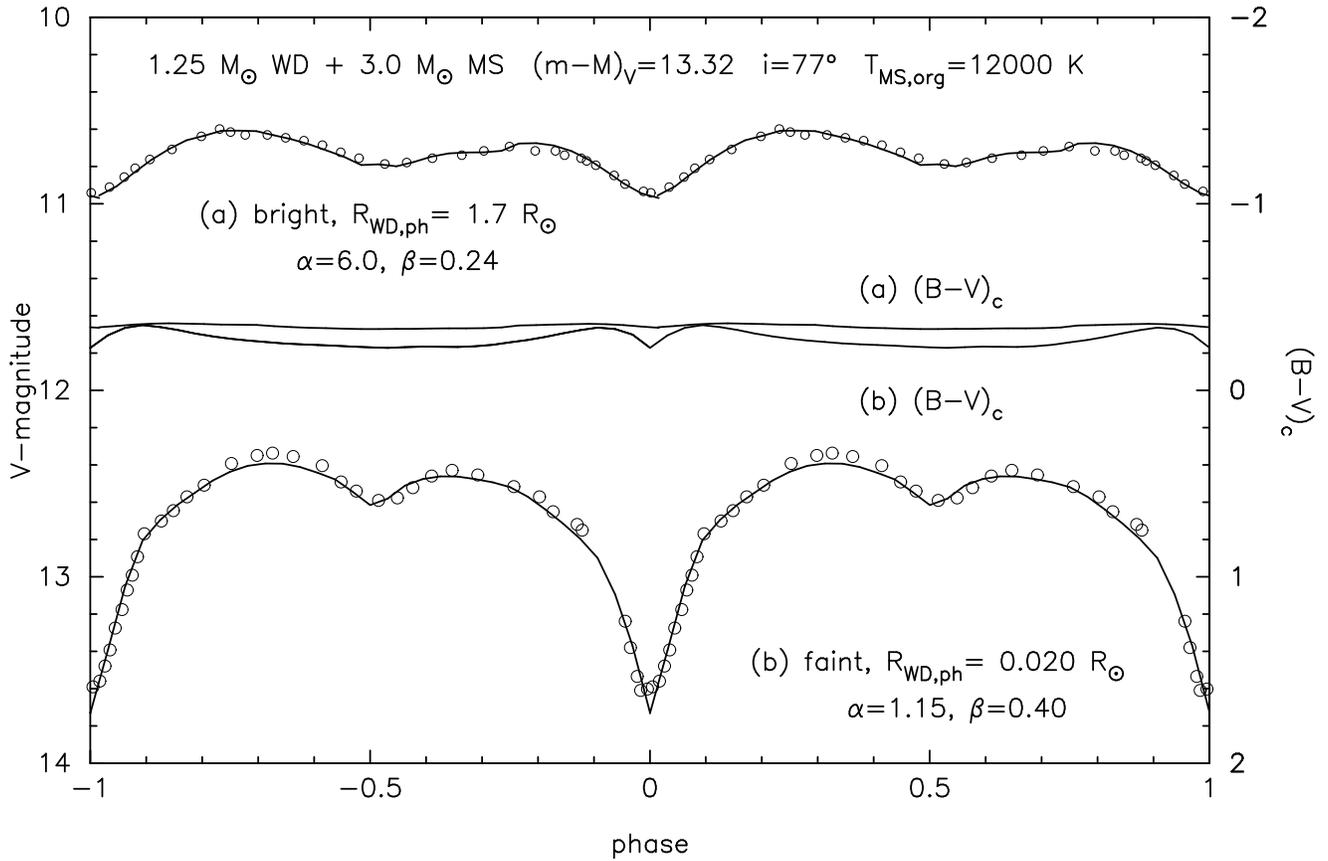}
\caption{
Calculated $V$ light curves and $B-V$ colors
are plotted against the binary phase 
(binary phase is repeated twice from $-1.0$ to $1.0$) together with
the observational points \citep{mad97, her65}.
Solid lines denote our calculated models.
Various system parameters are shown in the figure together with
the best fitted parameters.
(a) the bright (optical high) state.
(b) the faint (optical low) state.
\label{vmag_w125m30_orbital}}
\end{figure}

\begin{figure}
\plotone{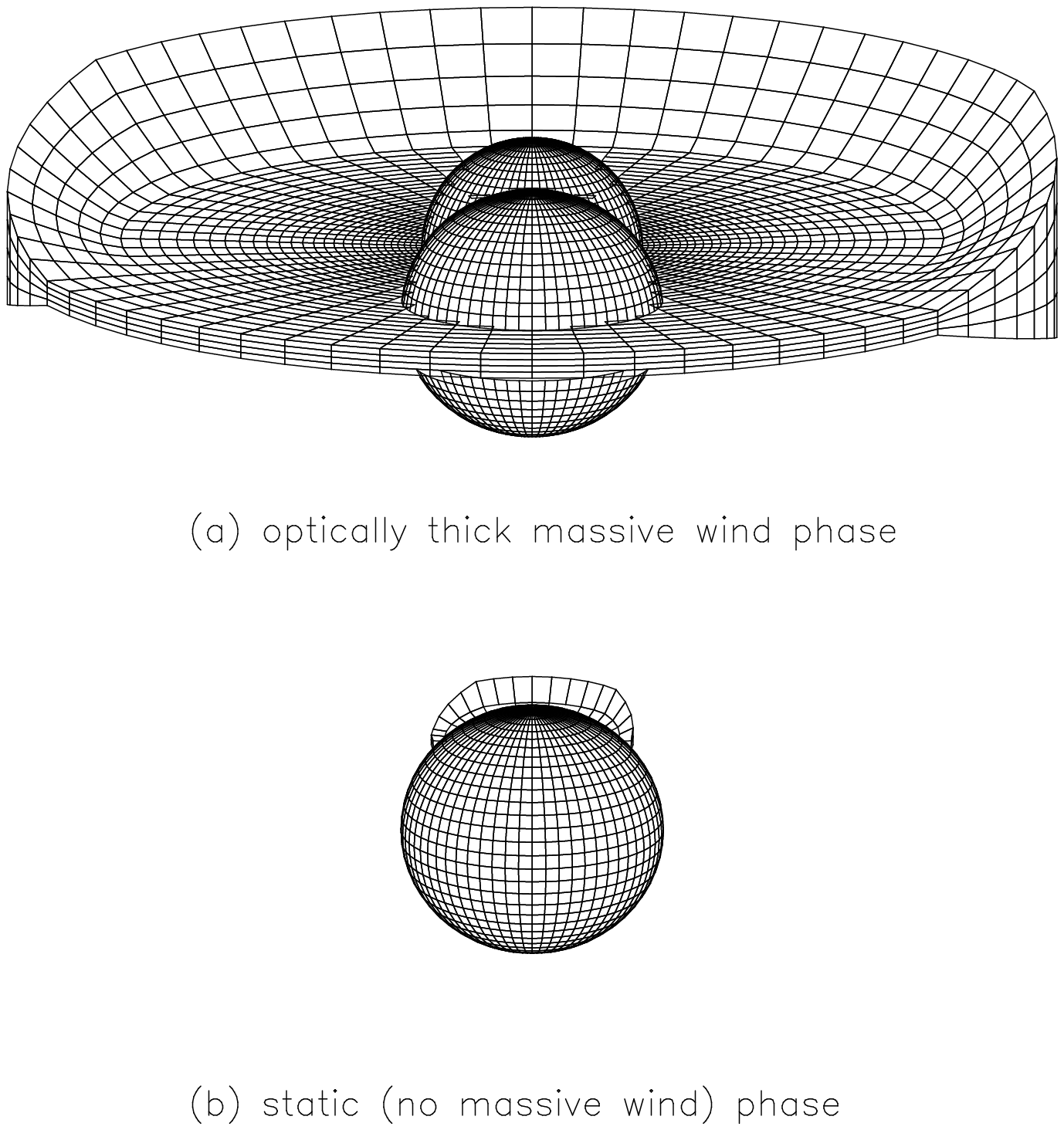}
\caption{
Same as Fig.\ref{w125m30configure80} but for the eclipse minimum
($\phi = 180\arcdeg$):
(a) in the massive wind phase (optical high/soft X-ray off)
and (b) in the static shell-burning (no massive wind) phase
after the wind stops
(optical low/soft X-ray on).  The WD surface is completely 
blocked by the MS companion but a part of the disk is seen
from the Earth.
\label{w125m30configure180}}
\end{figure}

\begin{figure}
\plotone{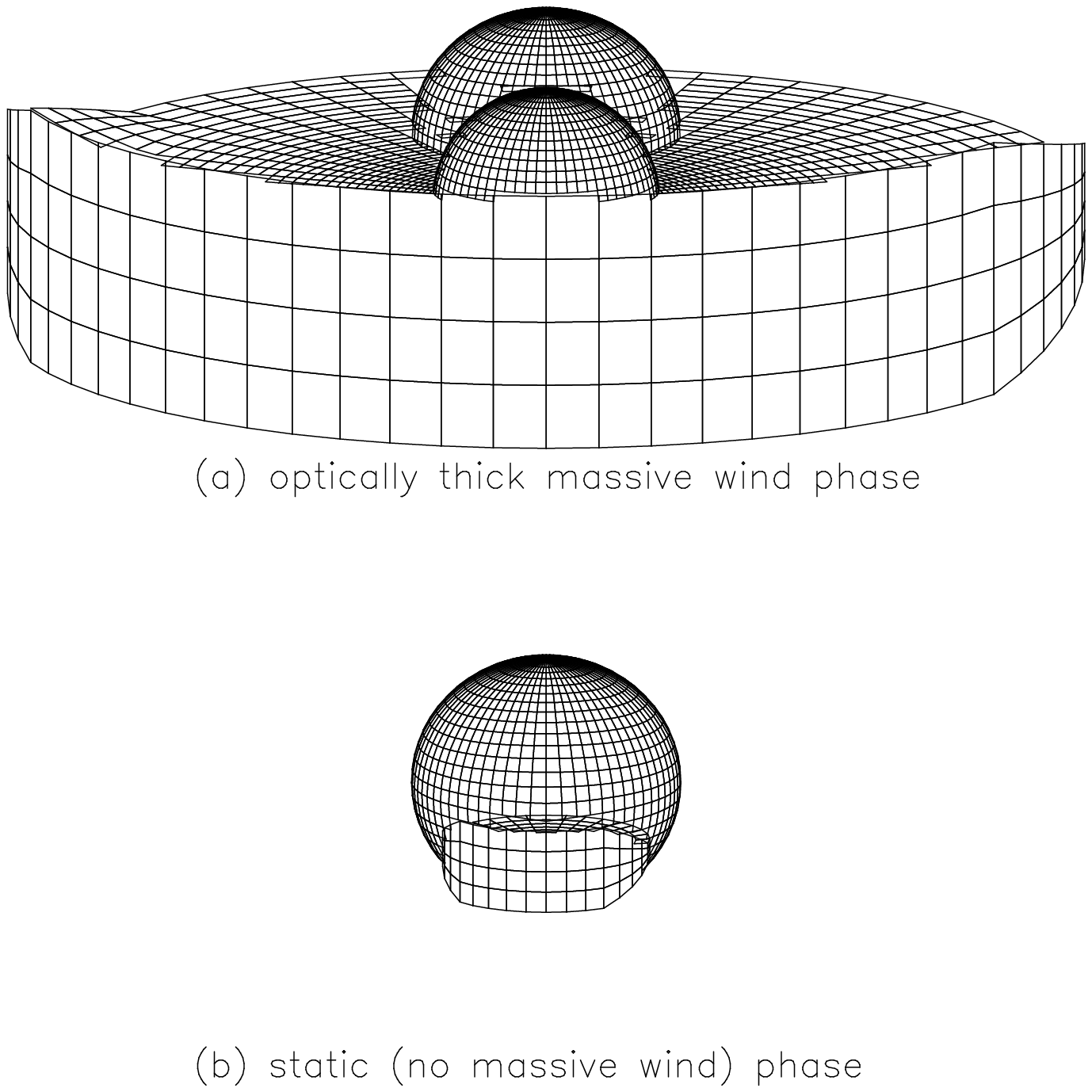}
\caption{
Same as Fig.\ref{w125m30configure80} but for the secondary minimum
($\phi = 0\arcdeg$):
(a) in the massive wind phase (optical high/soft X-ray off)
and (b) in the static shell-burning (no massive wind) phase
after the wind stops
(optical low/soft X-ray on).  The WD surface is blocked by
the edge of the disk.
\label{w125m30configure00}}
\end{figure}

     Our modeled light curves are shown 
in Figure \ref{vmag_w125m30_orbital}.
To reproduce the bright state, we must have a large size of the disk,
$\alpha \sim 6.0$, and an expanded WD envelope of 
$R_{\rm WD, ph} \sim 1.7~R_\sun$ 
($T_{\rm WD, ph} \sim 60,000$~K), as shown in Figure 
\ref{w125m30configure80}a.  For the optical faint state, 
we adopt a much smaller photospheric radius of 
$R_{\rm WD, ph} \sim 0.02~R_\sun$ and a smaller size of the disk, i.e., 
$\alpha= 1.15$, as shown in Figures \ref{w125m30configure80}b,
\ref{w125m30configure180}b, and  \ref{w125m30configure00}b.
The $1.25 ~M_\sun$ WD blows winds
when $R_{\rm WD, ph} \gtrsim 0.07 ~R_\sun$, so that these two
photospheric radii 
are consistent with the optically thick wind theory.
\par
     The photospheric radius of the WD ($R_{\rm WD, ph}= 1.7 ~R_\sun$) 
exceeds its critical Roche lobe ($R^*_1= 1.34 ~R_\sun$) but
smaller than that of the MS secondary ($R^*_2= R_2= 2.0 ~R_\sun$).
The velocity of the wind is as fast as $\sim 1000$~km~s$^{-1}$ 
or more at
the WD photosphere and much faster than the orbital velocity
($\sim 400$~km~s$^{-1}$).  As a result, the wind is almost free stream
near the photosphere and hardly affected by the rotation effect
(or the tidal effect) even if the photosphere exceeds the Roche lobe.
\par
     The velocities of winds calculated are as fast as 
$1000-1500$~km~s$^{-1}$ \citep[e.g.,][]{kat94h}
and the wind mass loss rate is
as large as $10^{-6} - 10^{-4} M_\sun$~yr$^{-1}$ 
\citep[e.g.,][]{hac01kb}.  
Observationally wind velocities are reported to be,
for example, $4000$~km~s$^{-1}$ during the bright 
state \citep[see, e.g.,][]{koc86, woo00}.
Such a massive and fast wind affects the surface of the accretion disk.  
Because of a large velocity difference between the wind and the disk 
surface, it certainly drives the Kelvin-Helmholtz instability
at the interface.  Only the surface layer of the disk is
dragged away outward with the velocity at least several hundred 
km~s$^{-1}$ like a free stream moving outward.
This surface free stream is optically thick
near the original disk but becomes optically thin somewhere
outside because of geometrical dilution effect.  We regard 
the transition place from optically thick to thin as the edge
of the extended disk (see Fig. \ref{opaque_disk}).
In this sense, we should understand that the outer rim shown in
Figure \ref{w125m30configure80}a is not the matter boundary but
a transition from optically thick to thin regions of the tenuous
disk surface flow.
\par
     It should be noted that a high density part of the disk 
(dark-hatched region in Fig. \ref{opaque_disk}) is hardly
changed by this relatively tenuous disk surface free stream
because the internal density
of the disk is much denser than that of the WD wind.
The wind mass loss rate reaches about $1 \times 10^{-5} M_\sun$~yr$^{-1}$
and its velocity is $\sim 1000-4000$~km~s$^{-1}$, so that the density of 
the wind is estimated from the continuity ($\dot M_{\rm wind} = 
4 \pi r^2 \rho v$) to be about $10^{-11}$g~cm$^{-3}$ at the distance 
of $3~R_\sun$ from the center of the WD.
On the other hand, the density of the accretion disk
is about $1 \times 10^{-2}$g~cm$^{-3}$ at the same radius. 
Here, we assume the standard accretion disk model \citep{sha73}
with a mass accretion rate of $\sim 1 \times 10^{-5} M_\sun$~yr$^{-1}$.
\par
     In the faint state, on the other hand, 
small WD radii and disk sizes are required as illustrated 
in Figure \ref{w125m30configure80}b.  As suggested by \citet{sma01},
the size of the disk exceeds the Roche lobe, that is,
$\alpha= 1.15$, as shown in Figures \ref{w125m30configure80}b,
\ref{w125m30configure180}b, and  \ref{w125m30configure00}b.
The edge of the disk
still contributes to the total light even in the eclipse minima
(see Fig. \ref{w125m30configure180}b) and, as a result, the eclipse is
not total but partial.
\par
      The orbital light curves both in the high and low states vary 
from night to night \citep[e.g.,][]{her65, mad97, loc97}.
These variations are attributed to the dynamical changes of
the disk edge shape, which are frequently
observed in supersoft X-ray sources.  In this sense, 
our obtained shapes of the disk both in the high and low states
are not rigid but dynamically variable.
\par
     The calculated color variations
are roughly consistent with the observations, 
if the color excess is $E(B-V) \sim 0.3$ \citep[e.g.,][]{mad97, 
pat98}, i.e., for the faintest state,
$(B-V)_o \sim 0.0$
against $(B-V)_c \sim -0.3$, and, for the brightest state, 
$(B-V)_o \sim -0.1$ 
against $(B-V)_c \sim -0.4$, where 
$(B-V)_c$ is the calculated color (see Fig. \ref{vmag_w125m30_orbital})
and $(B-V)_o$ the observed color 
\citep[taken from, e.g.,][]{her65, sim01, sma01}.
Then, the distance to V Sge is estimated to be $D = 3.0$~kpc
with $(m-M)_V= 13.32$ and $E(\bv)= 0.30$.

\placefigure{w125m30configure180}
\placefigure{w125m30configure00}

\section{VARIOUS ARGUMENTS FOR/AGAINST THE WHITE DWARF MODEL}
     The most serious criticisms on the WD model of V Sge have been
proposed by \citet{sma01}.
They excluded the model of a disk around a compact
primary component mainly because such a model cannot reproduce
the orbital light curves in the faint phase.  However, 
their conclusions are derived based on the assumptions 
that (1) the accretion
disk is geometrically thin, that (2) their irradiation effect
does not include the diffusion effect between the irradiated
and non-irradiated regions of the secondary surface, and
that (3) the size of the disk is limited within $0.9$ times 
the Roche lobe size.  As already shown in the previous 
subsection, we are able to reproduce the orbital light curves
if these three assumptions are relaxed in V Sge.
\par
     As for assumption (1), \citet{sch97} have shown that
an elevated edge of the disk reproduce the light curve
of the supersoft X-ray source CAL 87, which is very
similar to the faint state of V~Sge 
\citep[e.g.,][]{dia95, gre98, pat98, sim96a}.  As for assumption (2),
\citet{sch97} also showed that a diffusion
effect on the secondary surface is important to explain a wide 
primary eclipse (or a round shape just outside the primary 
eclipse).  The same round shape just outside the primary eclipse
was also observed in the recurrent nova CI Aql 2000 outburst
and such a round shape can be reproduced by the same
effect \citep{led03}.
\par
     As for assumption (3), the mass transfer rate becomes as
large as the Eddington limit for white dwarfs in the optical
low (faint) state, i.e.,
\begin{equation}
\dot M_{\rm Edd} \equiv {{4 \pi c G M_{\rm WD}} \over \kappa}
{{R_{\rm WD}} \over {G M_{\rm WD}}}
\sim 1 \times 10^{-5} M_\sun \mbox{~yr}^{-1},
\end{equation}
as will be shown in the following section.  In such high
accretion rates, a part of matter overflows the inner critical
Roche lobe.
However, the size of the disk, i.e., $\alpha = 1.15$, 
is still smaller than the outer critical Roche lobe.
\par
     \citet*{hoa96} found that the X-ray flux from V Sge 
shows modulation with periodicities approximately equal to
a half, a third, and a fourth of the orbital period but
not with the orbital period itself.  \citet{sma01} counted it
one of the evidences against the model in which the hotter component
is a source of soft X-rays.  
In our model, however, the hot component (white dwarf)
is blocked at least twice, i.e., by the edge of the disk and by 
the companion as seen in Figure \ref{w125m30configure80}. 
Furthermore, in some configurations of the disk like
in Figure \ref{w125m30configure80},
this occultation of the hot component by the disk edge
continues from binary phase $\sim 0.4$ to $\sim 0.7$, i.e., 
$\sim 0.3$ orbital periods, which may explain a modulation frequency
of a third of the orbital period.

\section{A LIMIT CYCLE MODEL OF V SAGITTAE}
     The long-term optical behavior of V Sge shows a quasi-regular
transition between high and low states \citep[e.g.,][for 
a recent summary of the long-term variations]{sim99}.
In this section, we reproduce the long-term optical variations
based on an optically thick wind model of mass-accreting white dwarfs
\citep{hac01kb, hkn96, hkn99, hknu99, kat83, kat94h}.

\subsection{Winds of mass-accreting WDs}
     An essential point of our wind model is that WDs blow a 
fast and massive wind when the mass accretion rate exceeds
the critical rate, i.e.,
\begin{equation}
\dot M_{\rm acc} > \dot M_{\rm cr} \approx 0.75 \times 10^{-6} 
\left( {{M_{\rm WD}} \over 
{M_\sun}} - 0.4 \right)~M_\sun {\rm ~yr}^{-1},
\label{critical_rate}
\end{equation}
for $X=0.7$ and $Z=0.02$ \citep[see, e.g.,][for other cases
of hydrogen content and metallicity]{hac01kb}.
The wind velocity is as fast as or faster than
$v_{\rm wind} \sim 1000$~km~s$^{-1}$
and the wind mass loss rate is approximately given by
\begin{equation}
\dot M_{\rm wind} \approx \dot M_{\rm acc} - \dot M_{\rm cr},
\end{equation}
and usually as massive as $\sim 1 \times 10^{-6} M_\sun$~yr$^{-1}$.

\subsection{Mass stripping effects}
     When massive winds collide with the surface of 
the companion, the surface is shock-heated and ablated in the wind. 
We estimate the shock-heating by assuming that the velocity component
normal to the surface is dissipated by shock and its kinetic energy
is converted into the thermal energy of the surface layer.
We regard that gas is ablated and lost from L3 point
(L3 is the outer Lagrange point near the MS companion) 
when the gas gets the same amount
of thermal energy as the difference of the Roche potentials between
the MS surface and L3 point. 
Then the mass stripping rate is given by
\begin{equation}
{{G M} \over {a}} \left( \phi_{\rm L3} - \phi_{\rm MS} \right) \cdot 
\dot M_{\rm strip} 
=  {1 \over 2} v^2 \dot M_{\rm wind} \cdot \eta_{\rm eff} \cdot g(q),
\label{mass_stripping_original}
\end{equation}
where $M=M_{\rm WD}+ M_{\rm MS}$, $a$ is the separation of the binary,
$\phi_{\rm MS}$ the Roche potential at the MS surface, 
$\phi_{\rm L3}$ means the Roche potential at L3 point
near the MS companion, both of which are normalized by $GM/a$,
$\eta_{\rm eff}$ the efficiency of conversion from kinetic energy 
to thermal energy by shock, 
$g(q)$ is the geometrical factor of the MS surface 
hit by the wind including the dissipation effect (only the normal 
component of the wind velocity is dissipated), and $g(q)$ is only 
a function of the mass ratio $q=M_{\rm MS}/M_{\rm WD}$ 
\citep*[see][for more details]{hkn99}.  Here we modified 
equation (21) of \citet{hkn99} to include the effect of Roche lobe
overflow from L3 point.  Then the stripping rate is estimated as
\begin{equation}
\dot M_{\rm strip} = c_1 \dot M_{\rm wind},
\label{mass_stripping_rate}
\end{equation}
where
\begin{equation}
c_1 \equiv {{\eta_{\rm eff} \cdot g(q)} \over
{\phi_{\rm L3} - \phi_{\rm MS}}}
\left({{v^2 a} \over {2 G M}} \right).
\label{wind_strip_coefficient}
\end{equation}
The efficiency of conversion is assumed to be $\eta_{\rm eff}=1$. 
Here, we assume further that $M_{\rm WD}= 1.25~M_\sun$, 
$M_{\rm MS}= 3.0~M_\sun$.  Then, we have the separation of
$a = 4.375 R_\sun$, the total mass of the binary $M= 4.25 M_\sun$,
and the difference of Roche potentials 
$\phi_{\rm L3} - \phi_{\rm MS} = 0.3$,
and $g(q)=0.03$ for the mass ratio 
$q= M_{\rm MS}/M_{\rm WD} \sim 3$ \citep{hkn99}.
Substituting these values to equation (\ref{wind_strip_coefficient}),
we have
\begin{equation}
c_1 \approx
0.1 \left( {v \over {600~{\rm km~s}^{-1}}} \right)^2 \sim  5,
\label{mass_stripping_rate_real}
\end{equation}
for the wind velocity of $v_{\rm wind}= 4000$~km~s$^{-1}$.
Here, the maximum wind velocity observed is  $4000$~km~s$^{-1}$ 
in the optical bright state \citep[e.g.,][]{koc86, woo00},
which is a similar wind velocity in RX~J0513.

\subsection{Mass transfer attenuated by winds}
     In such a wind phase, the net mass accretion rate to the WD, 
$\dot M_{\rm acc}$, is modified as 
\begin{equation}
\dot M_{\rm acc} = \left\{
        \begin{array}{@{\,}ll}
          0~(\mbox{~or~}\epsilon), 
& \quad \mbox{~for~} \dot M_{\rm MS} \le \dot M_{\rm strip} \cr
          \dot M_{\rm MS} - \dot M_{\rm strip}.
& \quad \mbox{~for~} \dot M_{\rm MS} > \dot M_{\rm strip}
        \end{array}
      \right.
\label{accretion_rate}
\end{equation}
Here we adopt a small value of 
$\epsilon= 1 \times 10^{-7} M_\sun$~yr$^{-1}$  
when $\dot M_{\rm MS} < \dot M_{\rm strip}$ because the mass
accretion to the WD does not stop abruptly but probably continues
at least for a draining time of the accretion disk.  We do not know
the exact draining time of the accretion disk after the mass 
transfer from the MS stops.  Alternatively, we just assume a small rate
of the mass accretion $\epsilon$ to mimic the draining of the disk
during that $\dot M_{\rm MS} < \dot M_{\rm strip}$.
\par
     To know when the rapid mass accretion resumes,
we monitor the level of flow as
\begin{equation}
{{d } \over {d t}} M_{\rm flow}= \dot M_{\rm MS} - \dot M_{\rm strip}
- \dot M_{\rm acc},
\label{on_secondary_surface}
\end{equation}
with the initial value of $M_{\rm flow}=0$
\citep[see Fig. 3 of][for more details]{hac03kb}.  
This corresponds to when the surface of the MS companion
fills its critical Roche lobe.
We regard that the mass transfer is going on when $M_{\rm flow} \ge 0$
(overflow level).
During that $\dot M_{\rm strip} > \dot M_{\rm MS}$, 
$M_{\rm flow}$ decreases to a large negative value.  Here,
we regard that the mass transfer stops when $M_{\rm flow} < 0$.
Then the mass transfer from the MS stops or its rate drops 
to $\epsilon$.  Once the mass accretion rate drops below 
the critical rate, i.e., $\dot M_{\rm acc} < \dot M_{\rm cr}$,
the wind gradually weakens and eventually stops.
Then, the stripping effect vanishes ($\dot M_{\rm strip}= 0$) 
and the level of flow ($M_{\rm flow} < 0$) begins to rise toward
$M_{\rm flow} = 0$.  We start again the rapid mass accretion
when the level reaches $M_{\rm flow}=0$.  This period is considered
to be a thermal recovering time of the envelope of the MS companion.
It should be noted
that the level remains to be $M_{\rm flow} = 0$ if a steady-state
is reached, i.e., $\dot M_{\rm MS} - \dot M_{\rm strip}
- \dot M_{\rm acc} = 0$.

\subsection{Time evolution of winds}
  Time evolution of the photospheric radius and temperature 
of the WD are calculated from Kato \& Hachisu's (1994) 
optically thick wind solutions.  It is written, 
after \citet{hac01kb}, as
\begin{equation}
{{d } \over {d t}} \Delta M_{\rm env} = \dot M_{\rm acc} 
- \dot M_{\rm nuc} - \dot M_{\rm wind},
\label{on_white_dwarf_surface}
\end{equation}
where $\Delta M_{\rm env}$ is the envelope mass,
$\dot M_{\rm acc}$ the mass accretion rate,
$\dot M_{\rm nuc}$ the decreasing rate by nuclear burning, and
$\dot M_{\rm wind}$ the wind mass loss rate of the WD.
The photospheric radius, temperature, and luminosity are given
as a function of the envelope mass.  When the photospheric radius
shrinks smaller than the binary size, the irradiation effects 
both of the accretion disk and of the companion star become important.  
We include these irradiation effects together with
a shadowing effect by the disk on the companion star
as already described in the preceding section
\citep*[see, e.g.,][]{hac01kb, hac03a, hac03kb}.

\subsection{Delay of mass transfer via accretion disk}
     An important thing is that equation (\ref{on_white_dwarf_surface})
stands on the WD but equations 
(\ref{mass_stripping_original})$-$(\ref{on_secondary_surface}) 
are on the MS.
Therefore, the mass accretion rate, $\dot M_{\rm acc}$, may not be
the same on each side at the specific time, $t$,
because it takes a viscous timescale of the accretion disk for gas
to reach the WD surface from the MS surface.  
Then, we have
\begin{equation}
\left. \dot M_{\rm acc}(t) \right|_{\rm WD} = \left. 
\dot M_{\rm acc}(t - t_{\rm vis})
\right|_{\rm MS}.
\end{equation}
Here we adopt 
a viscous timescale of
\begin{equation}
t_{\rm vis} = {{R_{\rm disk}^2} \over {\nu}} = {{R_{\rm disk}} \over
\alpha_{\rm vis} } {{R_{\rm disk}} \over {c_{\rm s} H}}
\sim 40 \left({{\alpha_{\rm vis}} \over {0.1} } \right)^{-1} \mbox{~days},
\label{viscous_timescale}
\end{equation}
where $\nu$ is the viscosity, $\alpha_{\rm vis}$ the $\alpha$ parameter
of \citet{sha73}, $c_{\rm s}$ the sound speed, and $H$ the vertical height
of the accretion disk.  We adopt $R_{\rm disk}= 1.4~R_\sun$ 
and $H/R_{\rm disk} \sim  0.1$ for
$T_{\rm disk} \sim 30000-50000$~K and $c_{\rm s} \sim 30$~km~s$^{-1}$.

\subsection{Model parameters}
     Our long-term light curves of V Sge are
calculated for a given set of parameters: 
(1) the white dwarf mass, $M_{\rm WD}$,
(2) the companion mass, $M_{\rm MS}$, and its surface temperature,
$T_{\rm MS, org}$,
(3) the mass transfer rate from the MS companion, $\dot M_{\rm MS}$,
(4) the viscous timescale of the accretion disk, $t_{\rm vis}$,
(5) the coefficient of stripping effect, $c_1$,
and (6) the disk parameters of $\alpha$ and $\beta$ and other
factors which determine the disk shape.  The parameters 
(1)$-$(5) are summarized in Table \ref{high_low_states}
except for $T_{\rm MS, org}= 10,000$, 12,000, and 13,500~K
for $M_{\rm MS}=2.5$, 3.0, and $3.5~M_\sun$, respectively. 
As for the disk shapes, we simply adopt 
the same disk shape parameters 
as shown in Figure \ref{w125m30configure80}. 
Typical long-term light curves are shown in Figures 
\ref{vmag1300m35_long_time_fit_vsge}-\ref{vmag1100m35_long_time_fit_vsge}
for various white dwarf masses, i.e., $M_{\rm WD}=1.3$, $1.25$, $1.2$, 
and $1.1~M_\sun$, respectively.

\placetable{high_low_states}
\placefigure{vmag1300m35_long_time_fit_vsge}
\placefigure{vmag1250m35_long_time_fit_vsge}
\placefigure{vmag1200m35_long_time_fit_vsge}
\placefigure{vmag1100m35_long_time_fit_vsge}

\section{NUMERICAL RESULTS OF LIMIT CYCLE MODEL}
\subsection{Template light curves \label{template_subsection}}
\begin{figure}
\plotone{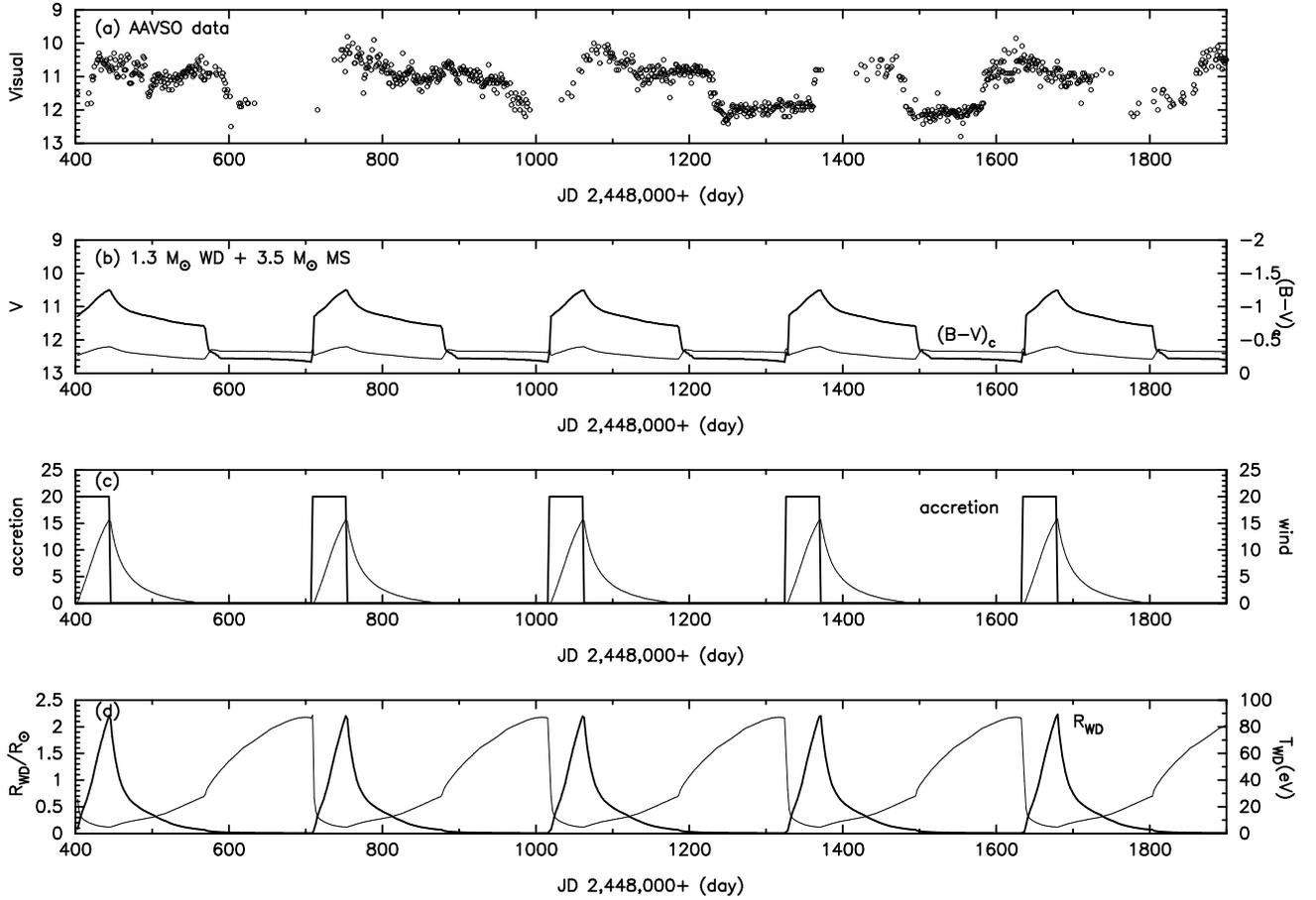}
\caption{
{\it Top panel}: (a) AAVSO visual data are plotted against time 
(JD 2,448,000$+$).  
{\it Second panel}: (b)
$V$-magnitude light curves (thick solid) 
of our V Sge model ($1.3~M_\sun$~WD and $3.5~M_\sun$~MS)
are plotted together with the calculated color $(B-V)_c$ (thin solid).
The calculated $V$-light curve is made by 
connecting the brightness at orbital phase 0.4.
The inclination angle is assumed to be $i = 77\arcdeg$.
{\it Third panel}: (c) 
Mass accretion rate to the WD ($\dot M_{\rm acc}$, thick solid)
and wind mass loss rate from the WD 
($\dot M_{\rm wind}$, thin solid), both
in units of $10^{-6} M_\sun$~yr$^{-1}$.
{\it Bottom panel}: (d) Photospheric radius of the WD envelope
in units of $R_\sun$ (thick solid) and surface temperature
of the WD envelope in units of eV (thin solid).
The model parameters are summarized in Table \ref{high_low_states}.
\label{vmag1300m35_long_time_fit_vsge}}
\end{figure}

\begin{figure}
\plotone{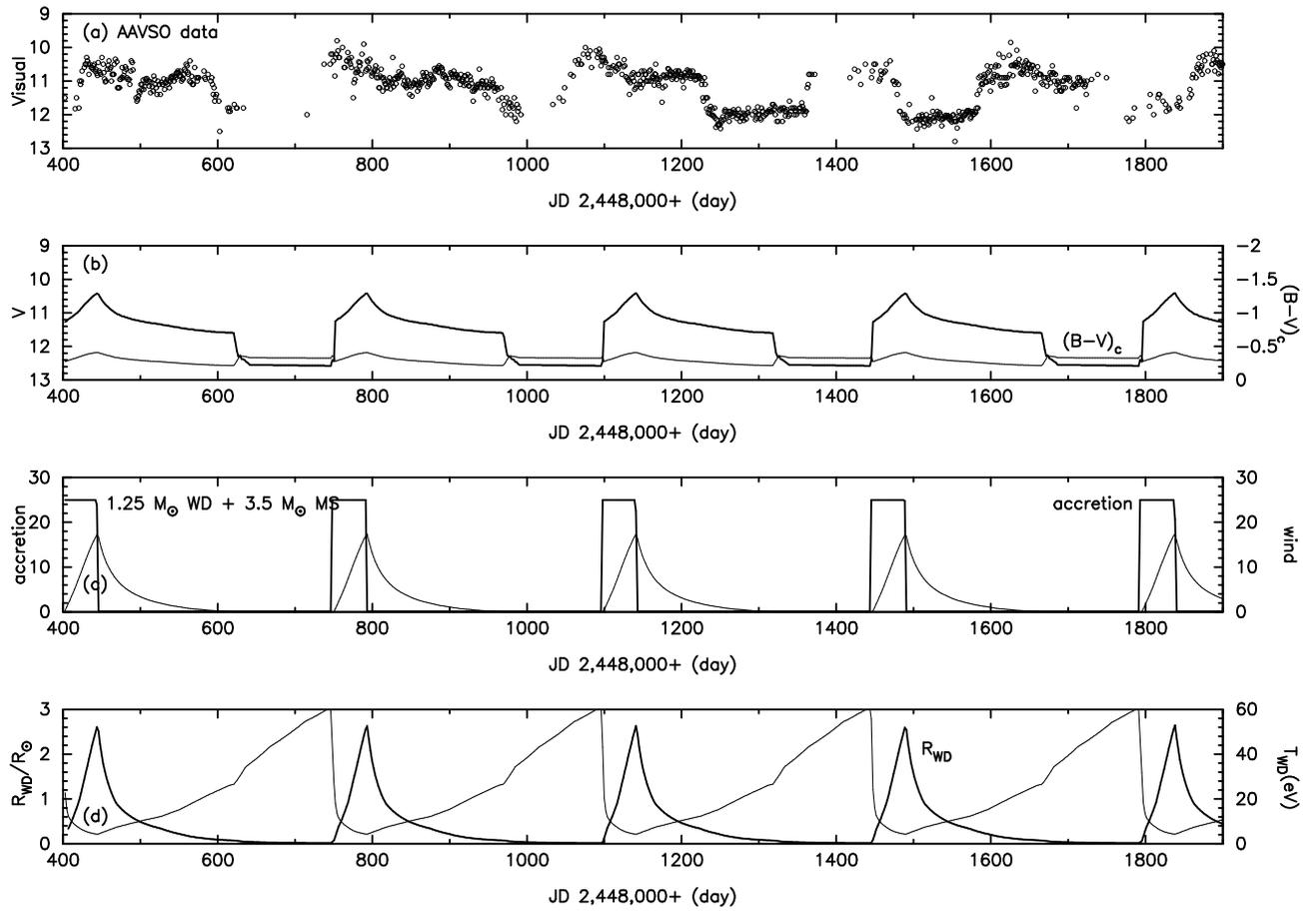}
\caption{
Same as Figure \ref{vmag1300m35_long_time_fit_vsge}, but for 
$1.25~M_\sun$~WD $+$ $3.5~M_\sun$~MS.
\label{vmag1250m35_long_time_fit_vsge}}
\end{figure}

\begin{figure}
\plotone{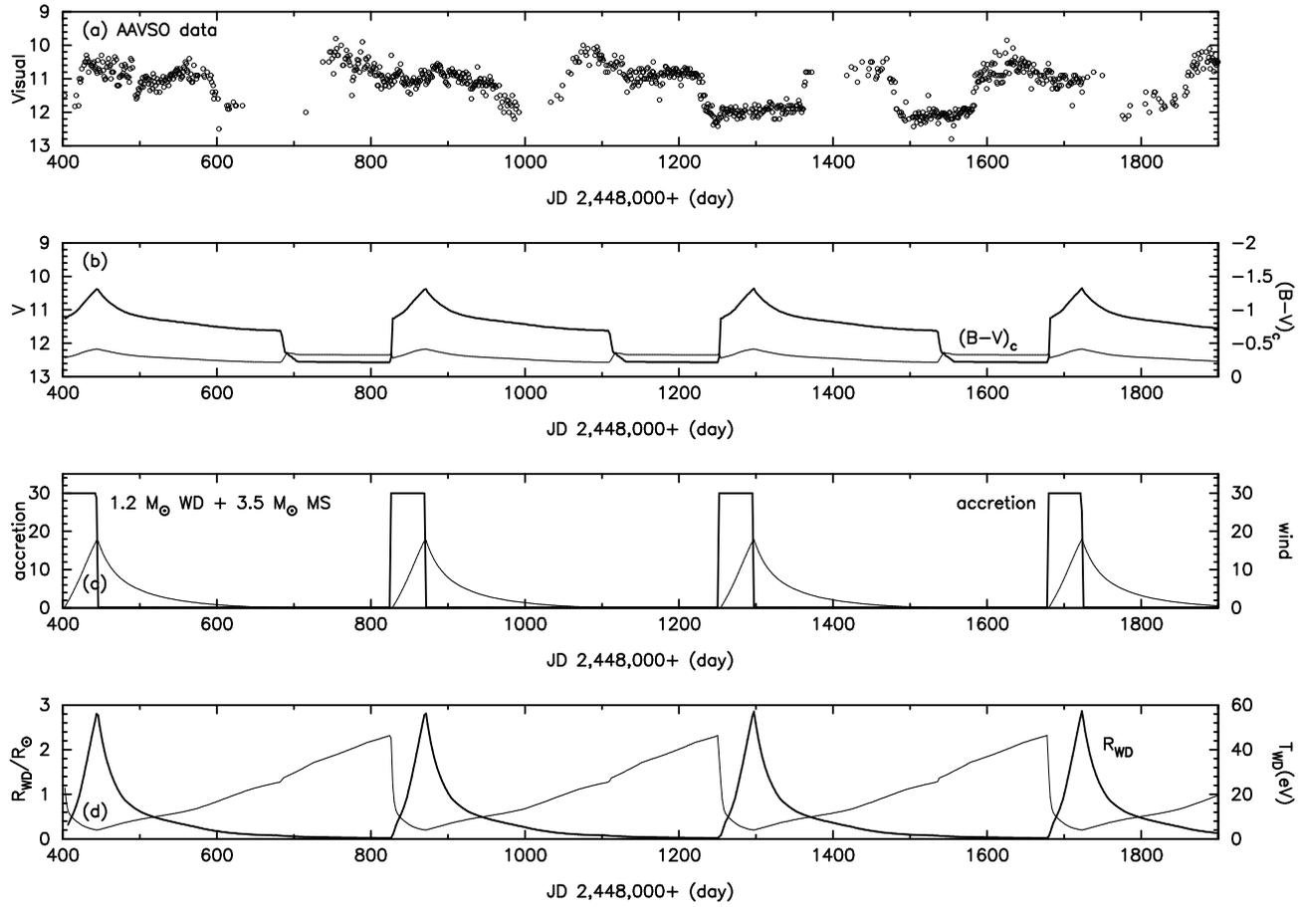}
\caption{
Same as Figure \ref{vmag1300m35_long_time_fit_vsge}, but for 
$1.2~M_\sun$~WD $+$ $3.5~M_\sun$~MS.
\label{vmag1200m35_long_time_fit_vsge}}
\end{figure}

\begin{figure}
\plotone{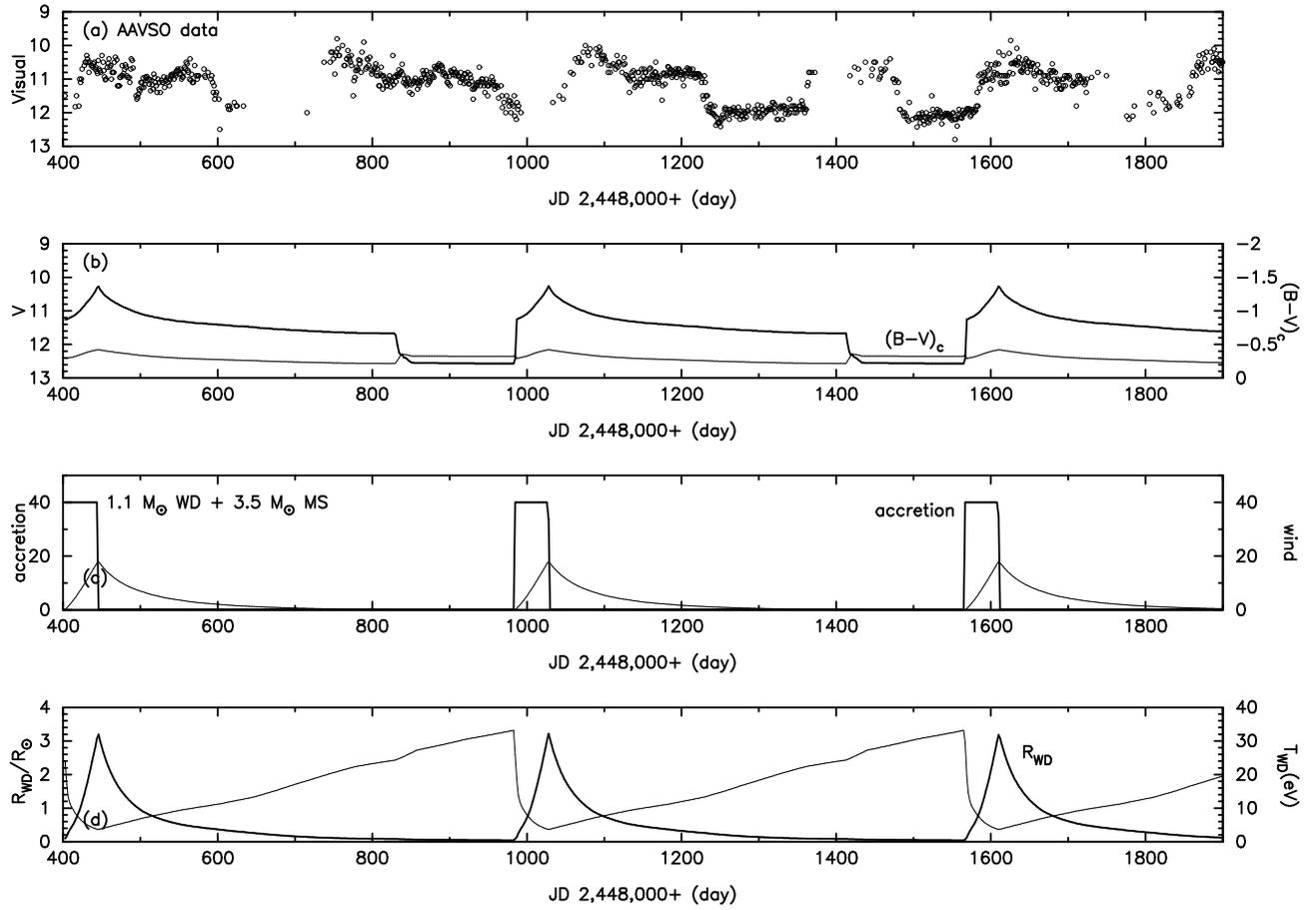}
\caption{
Same as Figure \ref{vmag1300m35_long_time_fit_vsge}, but for 
$1.1~M_\sun$~WD $+$ $3.5~M_\sun$~MS.
\label{vmag1100m35_long_time_fit_vsge}}
\end{figure}

     Now we show our long-term light curves that exhibit 
the transition between the optical high and low states.
The first case is a binary system
consisting of a $1.3~M_\sun$ WD and a $3.5~M_\sun$ MS companion.
Assuming $\dot M_{\rm MS} = 20 \times 10^{-6} M_\sun$~yr$^{-1}$,
$c_1 = 7.0$, and $t_{\rm vis}= 41.6$ days (i.e., 81 binary periods), 
we have followed the time evolution 
of our binary system with the initial envelope mass of
$\Delta M_{\rm env}= 3 \times 10^{-7} M_\sun$.
Figure \ref{vmag1300m35_long_time_fit_vsge} depicts 
the obtained $V$-light curve and color (second panel), 
accretion rate $(\dot M_{\rm acc}$ on the WD) 
and wind mass loss rate in units of $10^{-6}M_\sun$~yr$^{-1}$ (third
panel), radius ($R_\sun$) and temperature (eV) of the WD photosphere
(bottom panel) together with the AAVSO observation (top panel).
\par
     At the beginning of time-evolution calculations,
the mass of the WD envelope is small.
As the WD accretes matter and its envelope mass increases because of 
$\dot M_{\rm MS} = \dot M_{\rm acc} > \dot M_{\rm nuc}$.
When the envelope mass reaches the critical value,
$\Delta M_{\rm env} = 4.1 \times 10^{-7} M_\sun$, the WD photosphere
expands to $R_{\rm ph}= 0.07 ~R_\sun$ and begins to blow a wind.
Then the disk surface is blown in the wind and 
the disk size quickly expands from $\alpha= 1.15$ 
to $\alpha = 6.0$ in dynamical timescales.  Here we adopt seven days 
as the transition time, because the massive winds easily blow 
the surface layer of the disk in the wind.
The wind mass loss rate is about $\dot M_{\rm wind}= 5 \times 10^{-8}
M_\sun$~yr$^{-1}$ only one day after the wind starts, which is 
large enough to completely obscure soft X-rays 
\citep[see, e.g., discussion of][]{sou96}.
The occurrence of optically thick winds can naturally explain 
the very rapid emergence and decay of the supersoft X-rays
accompanied by the transition between the optical high and
low states \citep{gre98}.
\par
     The envelope mass continues to increase and the wind mass loss
rate also grows.  The WD photosphere continues to expand and 
this causes a gradual increase in the optical light curve.
The mass transfer from the MS is suppressed
when the wind mass loss rate reaches 
$\dot M_{\rm wind} = 2.9 \times 10^{-6} M_\sun$~yr$^{-1}$, because
$\dot M_{\rm strip} = 7 \dot M_{\rm wind} = \dot M_{\rm MS} =
20.0 \times 10^{-6} M_\sun$~yr$^{-1}$.
However, the mass accretion to the WD still continues at a high rate
and the wind mass loss rate reaches 
$\dot M_{\rm wind} = 15.0 \times 10^{-6} M_\sun$~yr$^{-1}$ 
as shown in Figure \ref{vmag1300m35_long_time_fit_vsge}
because the mass in the accretion disk
is drained to the WD for 41.6 days of $t_{\rm vis}$ time delay.
The mass accretion to the WD eventually stops and the mass of 
the WD envelope begins to decrease due to wind mass loss and nuclear
burning.
The WD photosphere gradually shrinks from $R_{\rm WD, ph}= 2.2
~R_\sun$ to $0.07 ~R_\sun$ during the wind phase.  This causes
the gradual decrease in the optical light curve until the end of 
high state.  When the mass of the WD envelope decreases to 
$\Delta M_{\rm env} = 4.1 \times 10^{-7} M_\sun$, the wind stops.
The duration of the wind phase is about 170 days as shown in
Figure \ref{vmag1300m35_long_time_fit_vsge} and summarized in
Table \ref{high_low_states}.
\par
     The WD photosphere begins to shrink quickly after 
the wind stops.  It takes 12 days to collapse from $0.07~R_\sun$
to $0.03~R_\sun$.  The photospheric temperature increases 
quickly from 30 eV to 40 eV during this period.
It takes 27 days to further contract from $0.03~R_\sun$
to $0.015~R_\sun$.  The photospheric temperature further increases 
from 40 eV to 60 eV during this period.
When the wind stops, the wind mass loss rate decreases 
from $\dot M_{\rm wind}= 1.5 \times 10^{-8} M_\sun$~yr$^{-1}$ to
zero in a day.  This means a sharp emergence of supersoft X-rays
in a day or so as observed in the LMC supersoft X-ray source
RX~J0513.9$-$6951 \citep{rei00}.
\par
     During the massive wind phase, the MS companion
is stripped off by the wind and its surface becomes smaller than
the critical Roche lobe, i.e., the level of mass flow,
$M_{\rm flow}$, in equation (\ref{on_secondary_surface})
goes to much below the level of zero.
When the wind stops, the surface of the MS companion is still
smaller than the Roche lobe.  It gradually recovers up 
and eventually reaches the level of zero.
It takes $\sim 90$ days in Figure \ref{vmag1300m35_long_time_fit_vsge}.
It takes a further 41.6 days until the mass accretion 
to the WD resumes, i.e., $t_{\rm vis}$ delay time.
It takes more 3 days that the WD envelope expands to blow a wind
again.  The total duration of low states amounts to
$90 + 40 + 3 \approx 130$~days.  Thus, the system repeats 
the cycle mentioned above, i.e., a 170 days optical high state 
and a 130 days optical low state.
\par
     The modeled system reaches a limit cycle after one cycle of high 
and low states.  Thus, the long-term light curve modulations are
roughly reproduced.  The time-averaged mass transfer rate 
to the WD is $3.0 \times 10^{-6} M_\sun$~yr$^{-1}$
and 80\% of the transferred
matter is blown in the wind, so that the MS loses its mass 
at the average rate
of $17.0 \times 10^{-6} M_\sun$~yr$^{-1}$ by stripping effects.
These values satisfy the conditions of $\dot M_{\rm MS}= 
\dot M_{\rm acc} + \dot M_{\rm strip}$ and $\dot M_{\rm strip} =
7 ~ \dot M_{\rm wind}$ in averaged values. 
\par
     \citet{rob97} divided the long-term optical brightness 
of \objectname{V Sge} into three levels,
i.e., the high ($V \lesssim 11.4$), 
intermediate ($11.4 \lesssim V \lesssim 12.0$), 
and low ($V \gtrsim 12.0$) states.
In this three-level classification of the brightness,
we attribute the phase of largely expanded WD photospheres
($R_{\rm WD, ph} \gtrsim 0.5~R_\sun$) 
to the real optical high ($V \lesssim 11.4$)
state.  When the WD photosphere expands as largely as
$\gtrsim 1~R_\sun$, it contributes more to the UV/optical light
compared with the disk irradiation.  The temperature of 
the WD photosphere is as high as $\sim 100,000$~K, while
the disk has lower temperatures of $10,000-50,000$~K.
  Thus, the $\bv$ color goes blueward as shown in Figures 
\ref{vmag1300m35_long_time_fit_vsge}$-
$\ref{vmag1100m35_long_time_fit_vsge}.
\par
     \citet*{loc97} detected radio emission from V~Sge and estimated 
the wind mass loss rate of
\begin{equation}
\dot M_{\rm wind} \sim 8 \times 10^{-6} 
\left({{D} \over {3 \mbox{~kpc}}} \right)^{3/2}
\left({{v_{\rm wind}} \over {1500 \mbox{~km~s}^{-1}}} \right)
 M_\sun \mbox{yr}^{-1}.
\end{equation}
The wind mass loss rate in our model (e.g., Fig.
\ref{vmag1300m35_long_time_fit_vsge})
reaches $\dot M_{\rm wind} \sim 15
\times 10^{-6} M_\sun$~yr$^{-1}$ 
and its averaged value during the wind phase
is as large as $\dot M_{\rm wind} \sim  5 \times 10^{-6} 
M_\sun$~yr$^{-1}$, which is very consistent with 
the radio observation.
\par
     Thus, we are able to reproduce the basic observational features 
(1)$-$(3) of V~Sge summarized in \S 1.  Furthermore, our model
naturally explains the other observational features:
(4) the WD photosphere continues to expand during the rapid mass accretion
phase and still goes on until $\sim 40$ days after the wind started.
This makes a visual magnitude peak of $\sim 0.5-1.0$ at the head 
of optical high states.  Large expansions ($\sim 2~R_\sun$)
of the WD photosphere 
are very consistent with the photospheric radius 
($\sim 1.7 ~R_\sun$) of the WD 
in the optical bright state
shown in Figure \ref{vmag_w125m30_orbital}.

\subsection{White dwarf mass}
     We have examined other WD masses,
i.e., $M_{\rm WD}= 1.25$, $1.2$, and $1.1~M_\sun$,
with the companion mass being kept to be $M_{\rm MS}=3.5 ~M_\sun$.
The results are shown in Figures
\ref{vmag1250m35_long_time_fit_vsge}, \ref{vmag1200m35_long_time_fit_vsge}
and \ref{vmag1100m35_long_time_fit_vsge}.
In general, more massive white dwarfs have a shorter duration
of the wind phase.  The durations of optical high and low states
are summarized in Table \ref{high_low_states} for various white dwarf  
masses and other parameters.  
\par
     Our guiding principle is that (1) we tune $c_1$ and
$\dot M_{\rm MS}$ 
to satisfy the duration of the low state (about 130 days),  
because we do not know the exact number of $c_1$, but it may
be around $3-10$ from equation (\ref{mass_stripping_rate_real}),
and that (2) we also tune $\dot M_{\rm MS}$ to reproduce 
the optical peak shape in the bright state.
Each tuned parameters are summarized in Table
\ref{high_low_states}.
\par
     For the case of $1.25~M_\sun$ WD $+$ $3.5~M_\sun$ MS,
we choose the parameters of $c_1 = 8.0$ and 
$\dot M_{\rm MS} = 25 \times 10^{-6} M_\sun$~yr$^{-1}$.
Then, the durations of the wind phase (optical high state)
and of the no wind phase (optical low state) are about 220 
and 130 days, respectively.  The observational light curve
shape of long-term variations is well reproduced by this
$1.25~M_\sun$ WD $+$ $3.5~M_\sun$ MS model.
\par
     When the rapid mass transfer from the MS companion stops,
we still keep a low rate of mass accretion to the WD, i.e.,
$\epsilon$ in equation (\ref{accretion_rate}).
To check the effect of $\epsilon$, 
we double the value of $\epsilon$ and calculate 
the long-term light curve.  The duration of the optical low
state becomes a bit shorter, 120 days, as listed in Table
\ref{high_low_states}.  On the other hand, 
the duration of the optical high state becomes a bit longer,
230 days, with the total period being almost not changed,
350 days.  We may conclude that the effect of $\epsilon$
is not so large as long as its value is
small compared with the rapid mass transfer rate.
This also examines the effect of the atmospheric mass
transfer, although the real number of the mass transfer
rate coming from the tenuous atmosphere of the MS companion
is not estimated, but it probably does not much exceed 
$1 \times 10^{-7} M_\sun$~yr$^{-1}$ \citep*[e.g.,][]{rit00}.
\par
     For the case of $1.2~M_\sun$ WD $+$ $3.5~M_\sun$ MS,
we have the parameters of $c_1 = 10.0$ and 
$\dot M_{\rm MS} = 30 \times 10^{-6} M_\sun$~yr$^{-1}$.
Then, the durations of the wind phase (optical high state)
and of the no wind phase (optical low state) are about 290 
and 140 days, respectively.  The calculated duration of
the optical high state (wind phase) is too long to be 
compatible with the observation.
\par
     For the case of $1.1~M_\sun$ WD $+$ $3.5~M_\sun$ MS,
we choose the parameters as $c_1 = 14.0$ and 
$\dot M_{\rm MS} = 40 \times 10^{-6} M_\sun$~yr$^{-1}$.
The durations of the optical high state
and of the optical low state are about 430 
and 150 days, respectively.  In this case, too, the calculated 
duration of the optical high state is much longer than
that of the observation.
\par
     When the durations of the low states are set to be similar 
to those for $M_{\rm WD}= 1.3~M_\sun$, the duration of high
states becomes much longer in the less massive white dwarfs.
From the theoretical light curve shapes and the durations of
optical high states, we may conclude that the WD mass in V~Sge
is somewhere between $1.3$ and $1.25 ~M_\sun$.
In what follows, we use $M_{\rm WD}= 1.25 ~M_\sun$ as already
adopted in equation (\ref{WD_mass}).

\begin{figure}
\plotone{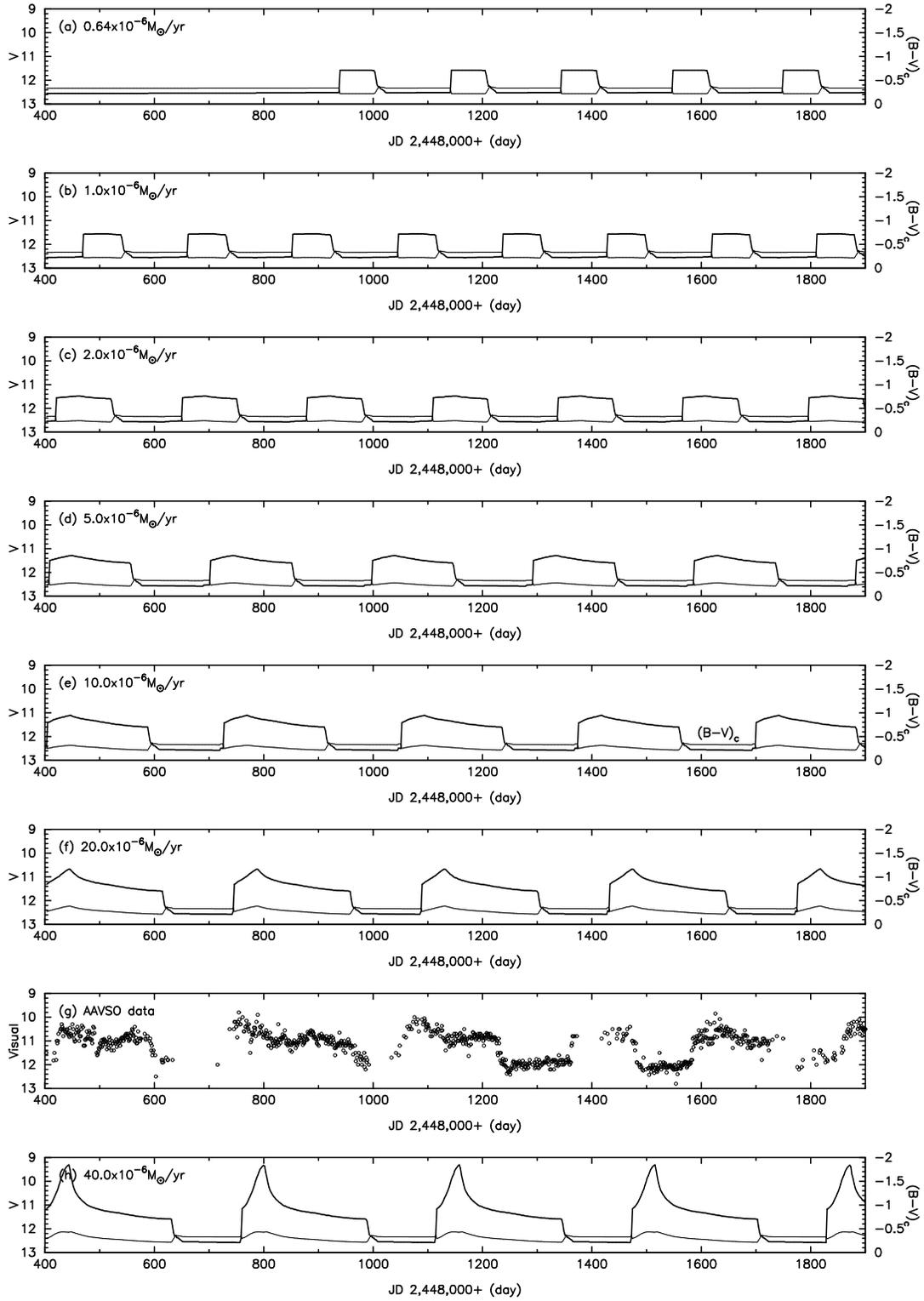}
\caption{
Calculated $V$-magnitude light curves (thick solid) of V~Sge
for various mass accretion rates
are plotted against time (JD 2,448,000$+$), together with 
the calculated color $(B-V)_c$ (thin solid).
The mass accretion rates are shown in each panel.
The other parameters are fixed and summarized in Table 
\ref{high_low_states}.  Here, the binary is
$1.25~M_\sun$ WD $+$ $3.5~M_\sun$ MS.
The inclination angle is assumed to be $i = 77\arcdeg$.
(g) AAVSO data of V~Sge taken from \citet{sim99} are also plotted.
\label{vmag1250m35_long_time_fit_vsge_acc}}
\end{figure}

\placefigure{vmag1250m35_long_time_fit_vsge_acc}

\subsection{Mass transfer rate of MS}
     The mass transfer rate of the MS companion depends on its
evolutionary stage and binary parameters
\citep[e.g.,][]{lih97, lan00}.  Since we do not know
the exact number of the mass transfer rate in V~Sge, 
we have examined various values of the original mass transfer
rate $\dot M_{\rm MS}$, i.e.,
$0.64$, $1.0$, $2.0$, $5.0$, $10.0$, $20.0$, 
$25.0$ (Fig. \ref{vmag1250m35_long_time_fit_vsge}), and $40.0$ 
in units of $10^{-6} M_\sun$~yr$^{-1}$ for the $1.25~M_\sun$ WD
$+$ $3.5~M_\sun$ MS.  Their results are summarized
in Table \ref{high_low_states} and 
the long-term light curves are shown 
in Figure \ref{vmag1250m35_long_time_fit_vsge_acc}.
Here, we fix the other parameters, i.e.,
$t_{\rm vis}= 41.6$ days and $c_1=8.0$.
The duration of optical high states becomes longer for the higher
mass transfer rates of $\dot M_{\rm MS}$.  On the other hand,
the duration of the optical low state hardly changes.
\par
     As the mass transfer rate of $\dot M_{\rm MS}$ increases,
an optical peak at the head of the high state becomes more prominent.
This is because the WD photosphere expands to as large as $1-2~R_\sun$
and both the irradiation effect by the disk and the WD brightness itself
increase and contribute much to the $V$-light.
From this peak luminosity, we adopt the mass transfer rate of
$\dot M_{\rm MS} = 25 \times 10^{-6} M_\sun$~yr$^{-1}$ for
the case of $1.25~M_\sun$ WD $+$ $3.5~M_\sun$ MS.

\begin{figure}
\plotone{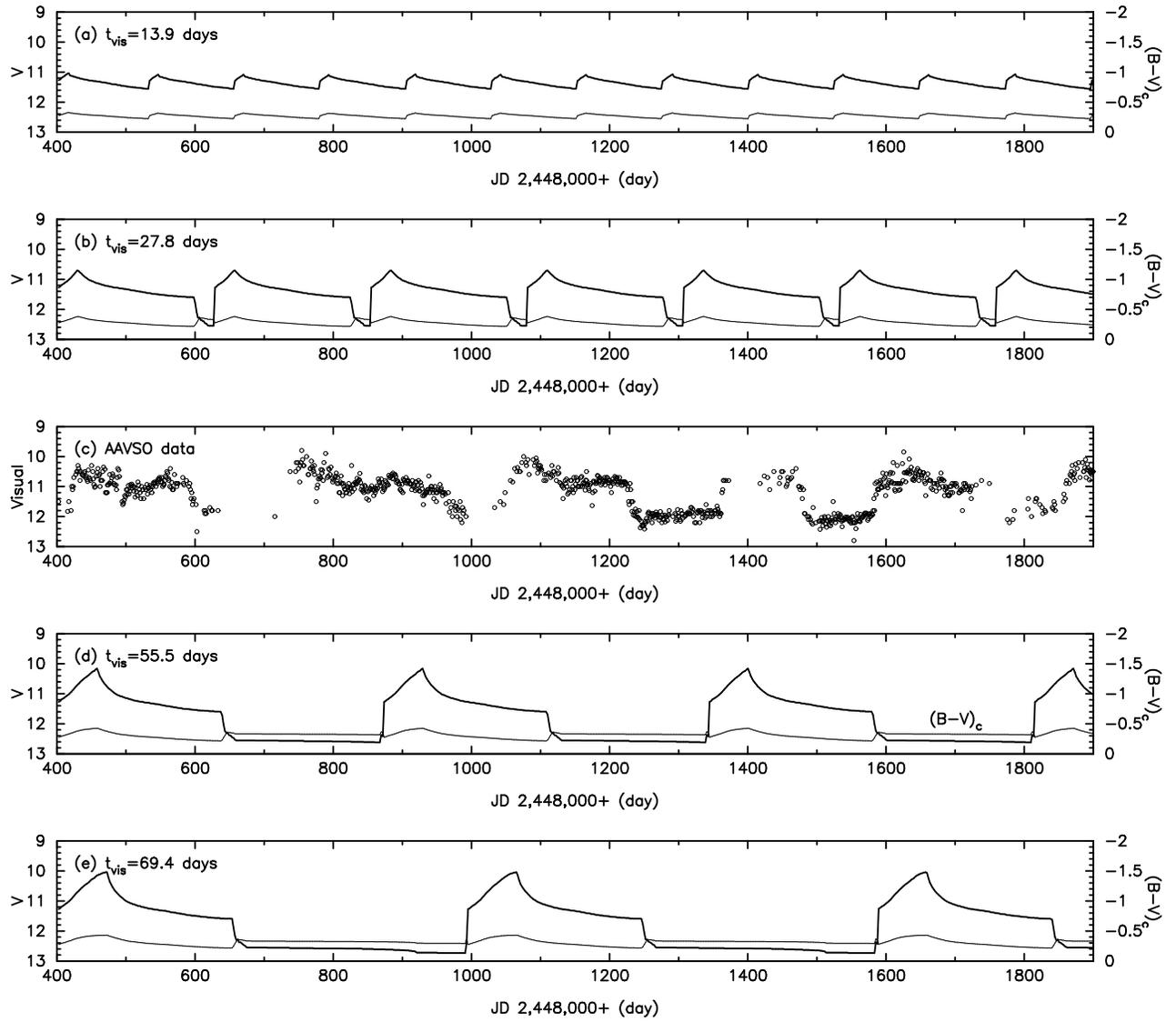}
\caption{
Same as Fig.\ref{vmag1250m35_long_time_fit_vsge_acc}, but for
various values of the delay by the accretion disk, $t_{\rm vis}$.
The value of $t_{\rm vis}$ is shown in each panel.
\label{vmag1250m35_long_time_fit_vsge_delay}}
\end{figure}

\placefigure{vmag1250m35_long_time_fit_vsge_delay}

\subsection{Effect of viscous timescale}
     The viscous timescale of accretion disks is not theoretically
determined yet.  Therefore, we examine the dependency of the value
of $t_{\rm vis}$ on our modeled light curves.  Here, we adopt
four other different values of 
$t_{\rm vis}= 13.9$ days (27 rotation periods),
$t_{\rm vis}= 27.8$ days (54 rotation periods),
$t_{\rm vis}= 55.5$ days (108 rotation periods),
and $t_{\rm vis}= 69.4$ days (135 rotation periods),
as summarized in Table \ref{high_low_states} and shown in Figure
\ref{vmag1250m35_long_time_fit_vsge_delay}.  The other parameters
are fixed, i.e., $\dot M_{\rm MS}= 25 \times 10^{-6} M_\sun$~yr$^{-1}$
and $c_1 = 8.0$ for the binary model of 
$1.25~M_\sun$ WD $+$ $3.5~M_\sun$ MS.
\par
     For the case of $t_{\rm vis}= 13.9$ days, we have no low states
because the rapid mass accretion resumes 13.9 days after the condition
$\dot M_{\rm MS} > \dot M_{\rm strip}$ is satisfied.  This time is
before the wind stops.
For the case of $t_{\rm vis}= 27.8$ days, we have a short duration
of low states, 30 days, much shorter than the observational ones.
\par
     On the other hand, for a relatively long timescale of 
$t_{\rm vis}= 55.5$ days, we have a rather long duration
of low states, 240 days, much longer than the observational ones.
This is because the rapid mass accretion still continues during 55.5 days
after the condition $\dot M_{\rm MS} < \dot M_{\rm strip}$ is fulfilled
and, as a result, a massive and fast wind strips off 
the surface of the MS companion
deeply, and it takes a much longer time to recover the level of zero
in equation (\ref{on_secondary_surface}).
For the case of $t_{\rm vis}= 69.4$ days, we have a very long
duration of low states, 340 days.  The steady hydrogen shell-burning
stops at 270 days in the low state, that is, 70 days before 
the optical high state starts.  
\par
     We have a reasonable duration of optical low states when
the viscous timescale is around $t_{\rm vis}= 40$ days, which 
is roughly corresponding to $\alpha_{\rm vis}= 0.1$ of \citet{sha73}
as seen in equation (\ref{viscous_timescale}).

\begin{figure}
\plotone{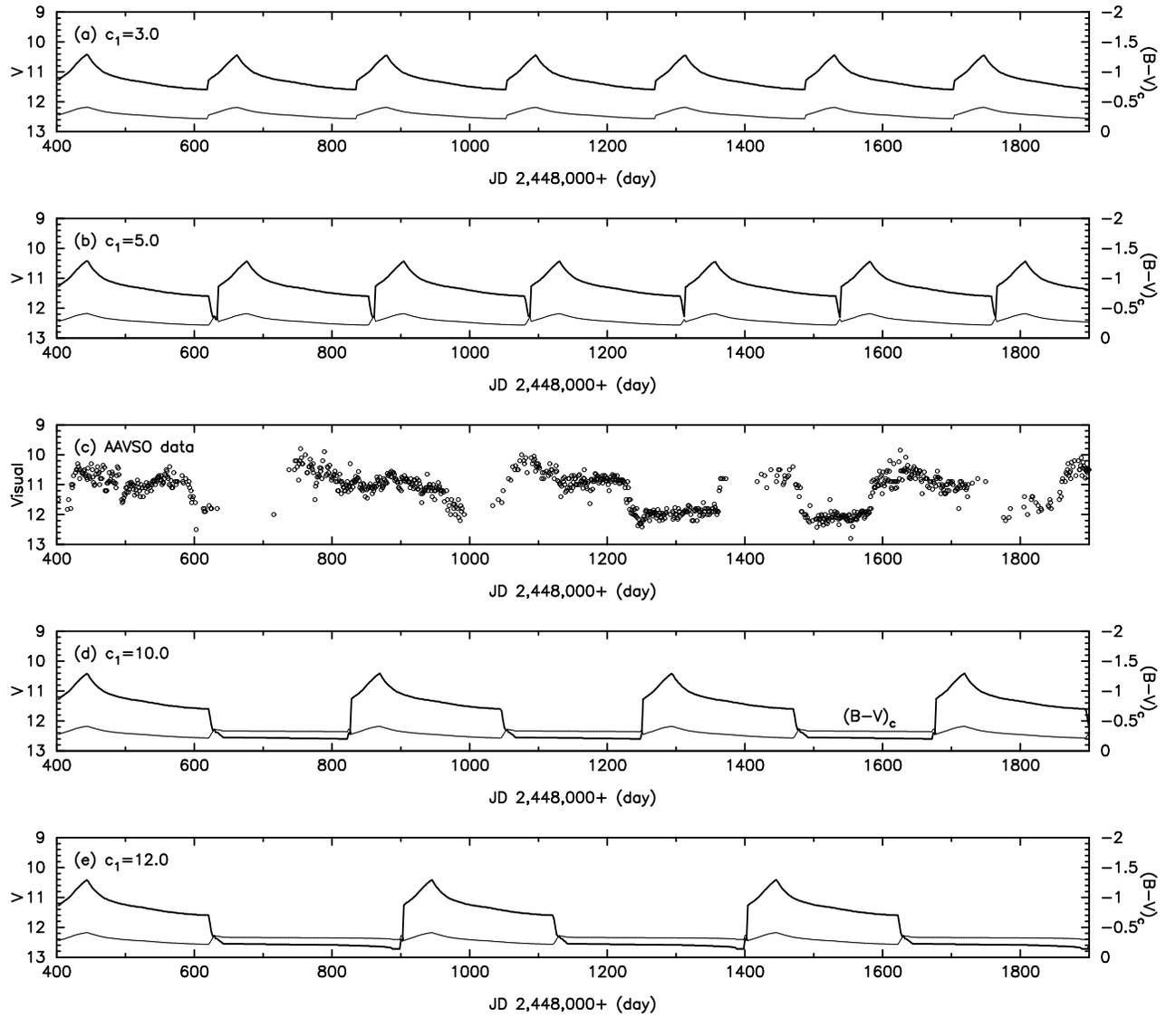}
\caption{
Same as Fig.\ref{vmag1250m35_long_time_fit_vsge_acc}, but for
various values of the stripping effect, $c_1$.
The value of $c_1$ is shown in each panel.
\label{vmag1250m35_long_time_fit_vsge_strip}}
\end{figure}

\placefigure{vmag1250m35_long_time_fit_vsge_strip}

\subsection{Mass stripping effect}
      It is a difficult work to accurately estimate the effect of
mass-stripping, that is, to determine the coefficient of $c_1$ in
equation (\ref{mass_stripping_rate}).  Therefore, we calculate
other four cases of $c_1$, i.e., $c_1=3.0$, $c_1=5.0$, 
$c_1=10.0$, and $c_1=12.0$ and examine how the light curves
depend on the coefficient.
It is obvious that strong modulation of the mass transfer rate
never occurs for the case of $c_1 \le 1$ because there is no case of 
$\dot M_{\rm MS} < \dot M_{\rm strip} = c_1 \dot M_{\rm wind}
(\le \dot M_{\rm wind} < \dot M_{\rm acc} < \dot M_{\rm MS})$.
Our calculated models and their parameters are listed 
in Table \ref{high_low_states}.
\par
     For the case of $c_1 = 3.0$ shown in Figure 
\ref{vmag1250m35_long_time_fit_vsge_strip},
we have no optical low states, mainly because 
the mass-stripping effect is small and, as a result,
the recovery time of the MS envelope is rather short.
     For the case of $c_1 = 5.0$,
the duration of optical low states is about 10 days, still 
because of small mass-stripping effect.
\par
     On the other hand, for the case of $c_1= 10.0$,
the mass-stripping effect increases and
the recovery time of the MS envelope becomes drastically longer.
If we further increase 
the value of $c_1= 12.0$, the duration of low states reaches
280 days, where the steady hydrogen shell-burning
stops at 270 days in the low state, i.e., 10 days before
the optical high state starts. 

\begin{figure}
\plotone{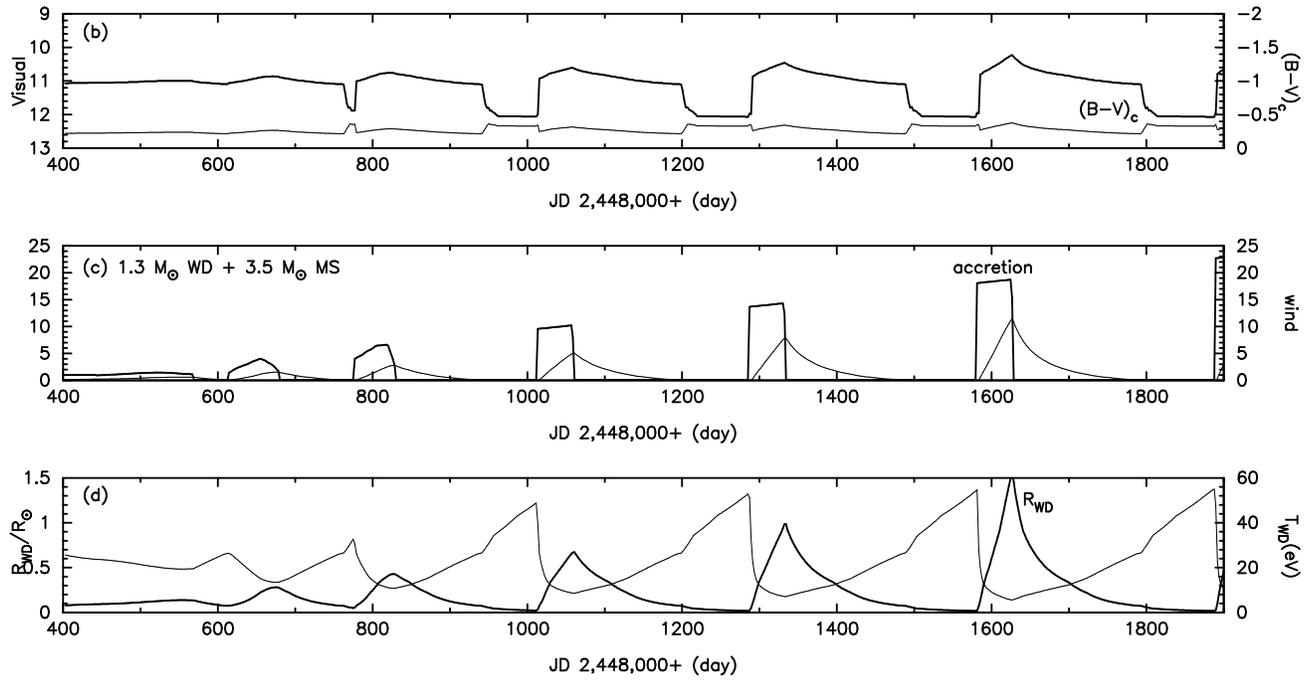}
\caption{
Transition from a flat segment to an active segment in the optical
light curve is calculated.  Both $c_1$ and $\dot M_{\rm MS}$
are gradually increasing with time.
The value of $c_1$ is going up from $c_1=1.0$ to $c_1=8.0$
and the mass transfer rate is also growing from 
$\dot M_{\rm MS}= 1 \times 10^{-6} M_\sun$~yr$^{-1}$ 
to $\dot M_{\rm MS}= 25 \times 10^{-6} M_\sun$~yr$^{-1}$,
both in 1200 days.
\label{vmag_long_time_fit_vsge_increase}}
\end{figure}

\placefigure{vmag_long_time_fit_vsge_increase}

\subsection{Flat segment to active segment}
     In the long-term light curve behavior, 
there are periods with/without strong optical modulations.
\citet{sim99} called them active/flat segments, respectively.
The active segments can be understood as a limit cycle of
the optical high and low states 
as already shown in the previous subsections.
Here, we interpret the nature of the flat segments.  
\citet{sim01} pointed out that the flat segments finished 
by another decrease into a real low state and further 
suggested that the intermediate (medium) state is
classified as the flat segment (see \S\ref{template_subsection}).
\par
     Our WD wind model provides a natural explanation of 
such behavior if the mass transfer rate
from the MS companion varies in time
together with the mass-stripping effect.
In the calculation shown
in Figure \ref{vmag_long_time_fit_vsge_increase},
we gradually increase $\dot M_{\rm MS}$ from 
$1 \times 10^{-6} M_\sun$~yr$^{-1}$ 
to $25 \times 10^{-6} M_\sun$~yr$^{-1}$ in 1200 days and,
at the same time, strengthen the mass-stripping effect 
from $c_1 = 1.0$ to 8.0.  This may imply that the wind velocity
(the mass-stripping effect) is relatively slow (small)
at the low wind mass-loss rate regime
but gradually increases proportionally to the wind mass-loss rate.
No modulations appear when the mass stripping effect is as small as
$\sim 1.0-1.5$ and the mass transfer rate is also as low as
$\sim 1 \times 10^{-6} M_\sun$~yr$^{-1}$.
This figure shows that modulation appears in the light curve
as both the mass transfer rate and the mass stripping effect
increase in time.

\placefigure{vmag_ms_mass_long_time_fit_vsge}

\subsection{Dependence on the secondary mass and the inclination angle}
\begin{figure}
\plotone{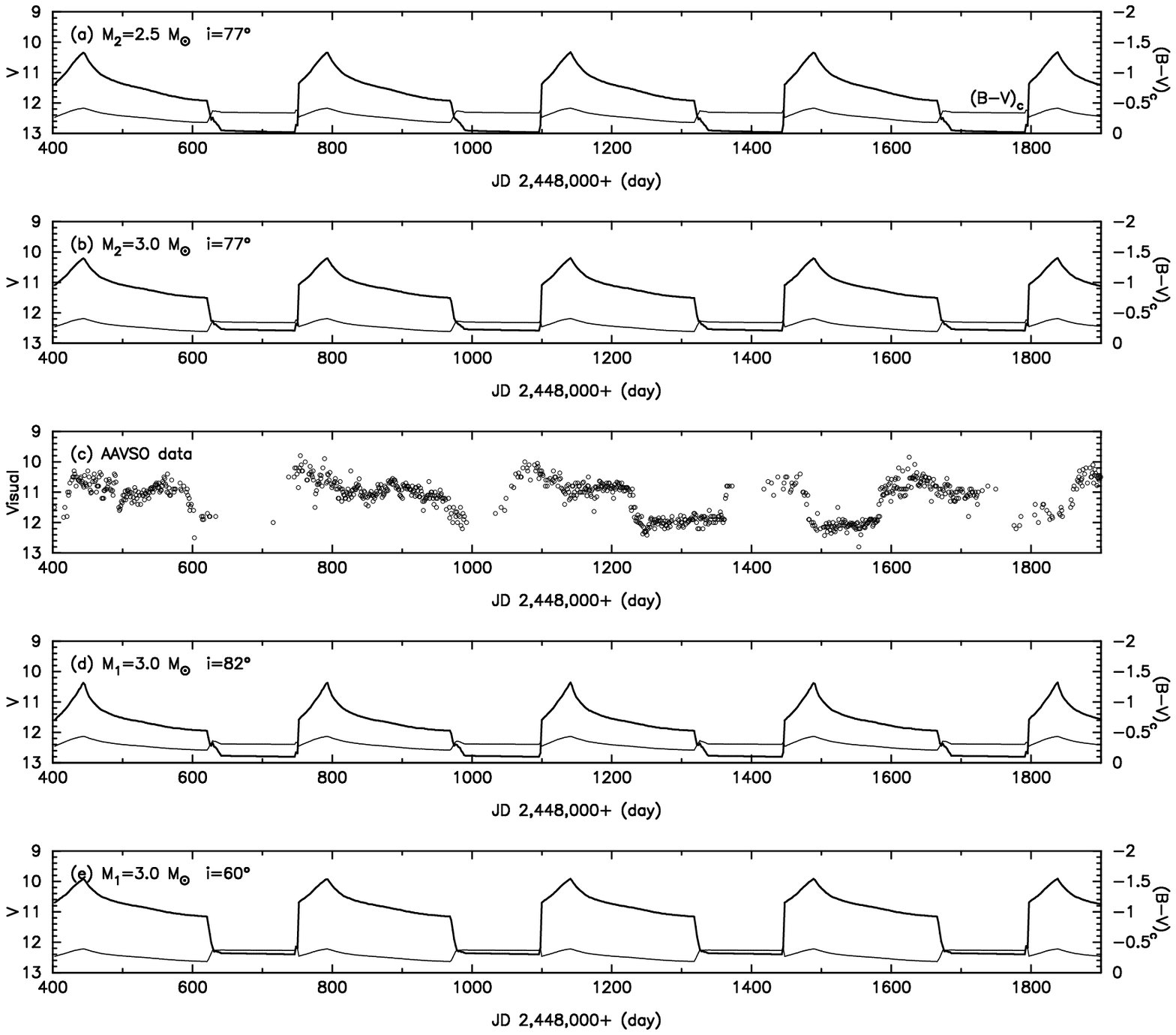}
\caption{
Same as Fig.\ref{vmag1250m35_long_time_fit_vsge_acc}, but for
other two cases of the companion mass and inclination angle, i.e.,
(a) $M_{\rm MS}= 2.5~M_\sun$ with the inclination angle of
$i= 77\arcdeg$, (b) $M_{\rm MS}= 3.0~M_\sun$ and $i= 77\arcdeg$,
(c)  AAVSO visual data taken from \citet{sim99},
(d)  $M_{\rm MS}= 3.0~M_\sun$ and $i= 82\arcdeg$, and
(e)  $M_{\rm MS}= 3.0~M_\sun$ and $i= 60\arcdeg$.
The shapes of $V$-light curves are not so different from each other,
although the depth of the low state becomes deeper for the lower
inclination angle ($i= 60\arcdeg$).
\label{vmag_ms_mass_long_time_fit_vsge}}
\end{figure}

     The long-term light curve behavior depends also on 
the other various parameters.   In this subsection, we examine
the long-term light curves by changing the other system parameters.
(1) The light curve also depends on the mass of the MS companion.
We have calculated two other masses of $M_{\rm MS}= 2.5~M_\sun$
and $3.0~M_\sun$ with the inclination angle of $i= 77\arcdeg$.
The results are shown in Figure \ref{vmag_ms_mass_long_time_fit_vsge}.
The overall light curve features are not changed but the brightness
becomes fainter in the less massive companion masses.
(2) We have also checked the dependency of the light curve on the 
inclination angle, i.e., $i= 82\arcdeg$ and $i= 60\arcdeg$ as
shown in Figure \ref{vmag_ms_mass_long_time_fit_vsge}.
The depth of the optical low state becomes much deeper 
at the lower inclination angles, that is, much deeper at
$i= 60\arcdeg$ than at $i= 80\arcdeg$.

\placefigure{he2_line_high}
\placefigure{he2_line_config}

\section{DISCUSSION}
\subsection{The \ion{He}{2} 4686 emission line in the high state} 
\begin{figure}
\plotone{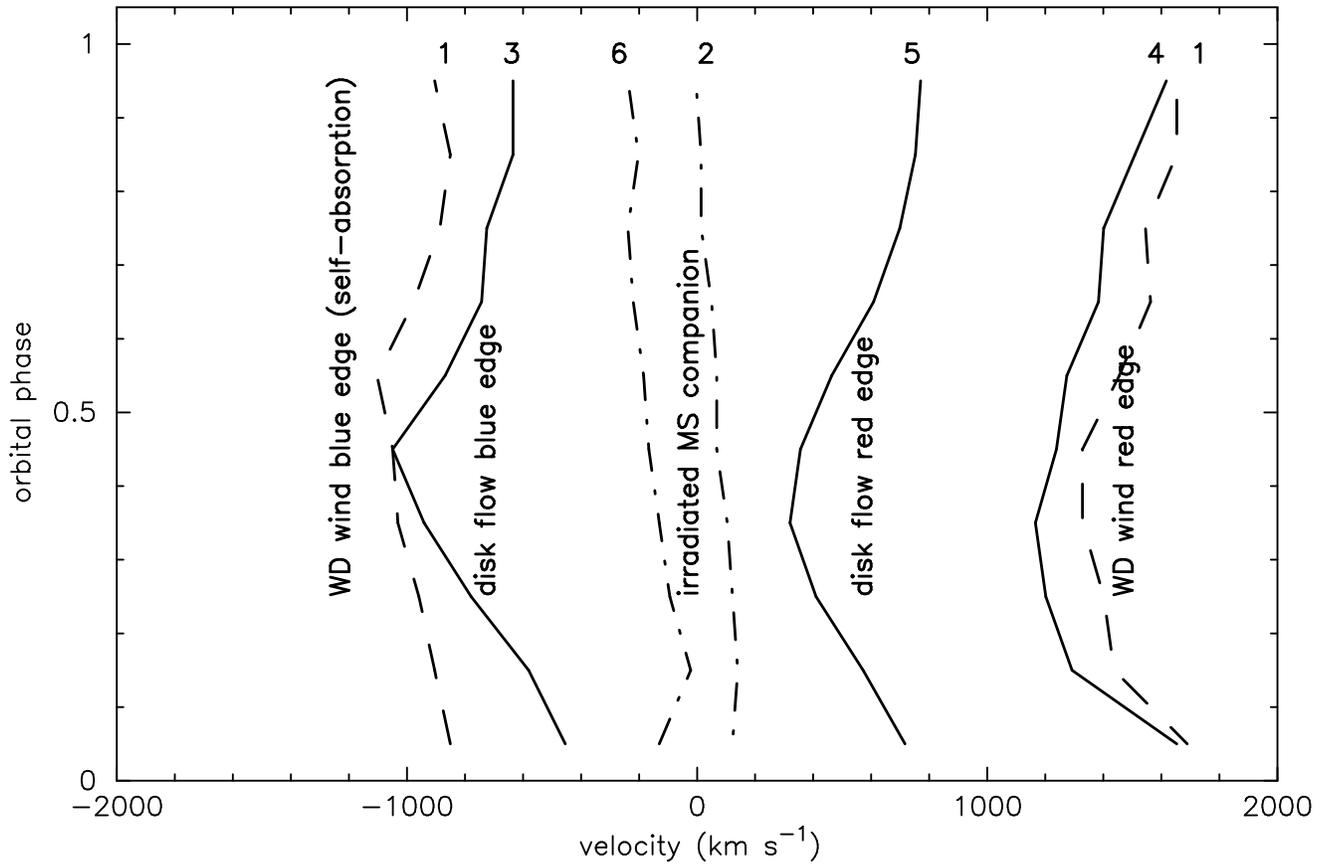}
\caption{
Velocities for \ion{He}{2} 4686 emission components 
are illustrated against the orbital phase.
Attached numbers denote the \ion{He}{2} 4686 emission components 
defined by \citet{loc98dt}.  These velocities are compiled 
from Fig. 1 of \citet{loc98} and Figs. 1 and 2 of \citet{woo97}.
See text for more details.
\label{he2_line_high}}
\end{figure}

\begin{figure}
\plotone{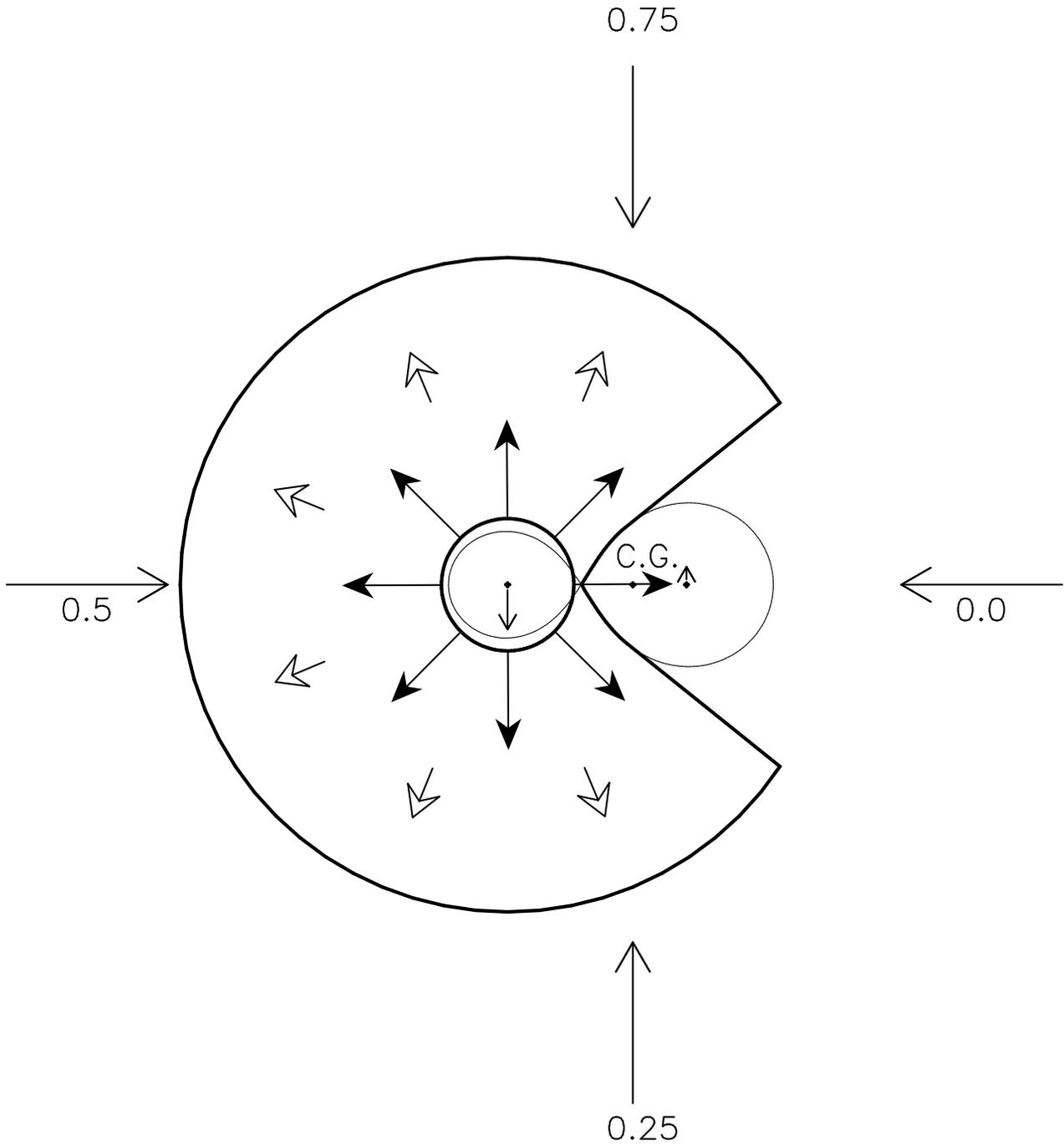}
\caption{
Line formation regions of \ion{He}{2} $\lambda$4686 in the 
optical high sates are schematically illustrated.
The WD (left component) blows a massive wind,
the velocity of which is as fast as $2000-2500$~km~s$^{-1}$ (long 
filled arrows).
The surface layer of the disk is dragged outward and
its surface extends up to several times the Roche lobe size.
Then, the disk surface is accelerated to 1000~km~s$^{-1}$ (short 
outlined arrows).   
Motions of binary components are denoted by open arrows, which are
starting from the center of each component.
The binary orbital phase is also denoted by outer open arrows.
\label{he2_line_config}}
\end{figure}

     Phase-resolved spectra in the optical high state 
have been observed
by several groups \citep*{gie98, loc98, rob97, woo97}.
\citet{loc98dt} presented high-quality, phase-resolved spectra
of \ion{He}{2} $\lambda$4686 and H$\beta$ emission lines
in the optical high state \citep[see also,][]{loc98, pat98, woo97}.
He showed six kinematic 
components in the \ion{He}{2} $\lambda$4686 emission line.
\citet{gie98} summarized them as follows: \\
(1) A broad, flat-topped component with steep edges that
extend further to the red than the blue.  This broad feature
also dominates H$\alpha$, and double-Gaussian fittings of
the extreme wind radial velocity variations are similar
in both \ion{He}{2} and H$\alpha$ \citep{gie98, rob97}.
The phase-resolved behaviors of these two sharp edges, which are
attached by number ``1'', are
shown in Figure \ref{he2_line_high}.  The left dashed-line 
corresponds to the blue edge while the right dashed-line
does to the red edge. \\
(2) A central, narrow, nebular line.  This component appears
stationary in both \ion{He}{2} and H$\alpha$.
The H$\alpha$ profiles appear double-peaked in both \citet{gie98}
and \citet{rob97}, because of the presence of both
\ion{He}{2} $\lambda$6560 and H$\alpha$.  The H$\alpha$ component
indicates an outflow with a velocity 
of $\sim 200$~km~s$^{-1}$.  The phase-resolved behavior of this
line is also shown in Figure \ref{he2_line_high}, to which 
number ``2'' is attached. \\
(3) An outer blue peak that accelerates blueward after 
primary eclipse and attains minimum velocity near the 
extreme blue wind just prior to phase 0.5, to which number ``3''
is attached.
This feature is also seen in H$\alpha$. \\
(4) An outer red peak, to which number ``4'' is attached,
best seen near phase 0.7, that
moves along with the red sharp edge of the broad component 1.\\
(5) An inner red peak that appears near phase 0.6, 
to which number ``5'' is attached, moves
redward from line center, and then disappears again near phase 0.1.
This feature is seen in H$\alpha$. \\
(6) An inner blue peak that emerges from the central nebular line 
to follow the motion of star 1 (primary component),
to which number ``6'' is attached.
This is seen as a slight blueward extension of the H$\alpha$
peak in the radial velocity range 0 to $+200$~km~s$^{-1}$
between phase 0.2 and 0.5.
\par
     With the above features, we presume three line 
formation regions of the \ion{He}{2} $\lambda$4686
in our WD optically thick wind model:
the first one is the WD massive wind, 
the second one is the irradiated surface of the disk, and
the third one is the irradiated side of the MS companion and
the matter which is stripped off from the MS companion.
The first and second regions are denoted by filled arrows
and short outlined arrows, respectively, 
in Figure \ref{he2_line_config}.  
\par
     The red sharp edge (component 1
and 4) almost follows the motion of star 1 (the primary WD)
with $K_1 \sim 250$~km~s$^{-1}$ but red-shifted 
by $\sim +1500$~km~s$^{-1}$.
It suggests that this component is originated from the WD wind
with the velocity of $\sim 1500$~km~s$^{-1}$.  The sharp blue edge of
emission component 1 is not consistent with this WD wind model
because it does not show a sinusoidal feature of 
$K_1 \sim 250-300$~km~s$^{-1}$.
This feature may be explained
by the effect of self-absorption by the WD wind itself, although
it is heavily contaminated by the \ion{C}{3}$+$\ion{N}{3} complex
in the blue edge over $\sim -1000$~km~s$^{-1}$.
The spectra of \citet{loc98} and \citet{pat98} certainly show
a weak feature of the \ion{C}{3}$+$\ion{N}{3} complex.
\par
     Emission components 3 and 5 also have a similar 
$K_1 \sim 250-300$~km~s$^{-1}$ and follow the motion 
of star 1 with a smaller red shift of 
$\sim +600$~km~s$^{-1}$ for component 5 and a blue shift of
$\sim -600$~km~s$^{-1}$ for component 3.
This suggests the velocity of 
the disk surface flow as fast as $\sim 600$~km~s$^{-1}$
(see also Fig. \ref{opaque_disk}).
\par
   For H$\alpha$ emission, \citet{gie98} obtained a radial velocity
curve of $K_1 \sim 380$~km~s$^{-1}$ by double-Gaussian fitting
for the inner flat-topped structure with $\sigma = 1000$~km~s$^{-1}$.
This implies a bit faster disk surface outflow, which is 
reasonably explained if the line formation region of
H$\alpha$ is further out than that of \ion{He}{2} 4686 and
the disk surface flow there is more accelerated. 
\par
     The central emission components 2 and 6 (dash-dotted lines
in Fig. \ref{he2_line_high}) show anti-phase motions
of the other components.
This feature indicates that components 2 and 6 are
originating from star 2.  As we have already discussed,
the MS companion (star 2) loses matter stripped by the WD wind
in the optical high state.  This gas outflow is much slower
compared with the WD wind and the disk surface flow, may
be as slow as $\sim 100-200$~km~s$^{-1}$.

\placefigure{pole_on_view}

\subsection{The \ion{He}{2} 1640 emission line in the low state} 
     The main features of the observed \ion{He}{2} 1640 
spectra in the optical low state \citep{woo00} are
summarized as follows: (1) At phase zero the line is roughly
symmetrical about its center although an excess in the red peak
is seen.  (2) As the binary phase increases, the emission on 
the blue side of the line increases until at phase $\sim 0.3$
with emission out to a velocity of $\sim -1500$~km~s$^{-1}$.
At this phase a sharper red edge appears at $\sim 300$~km~s$^{-1}$.
(3) At phase 0.4 the line is roughly symmetric again.
(4) Then, from phase 0.5 to 0.8, the situation is opposite
to phase from 0.1 to 0.4, with the extended emission now
occurring on the red side out to $\sim 1600$~km~s$^{-1}$ and
the sharper blue edge at $\sim -200$~km~s$^{-1}$.
\par
\begin{figure}
\plotone{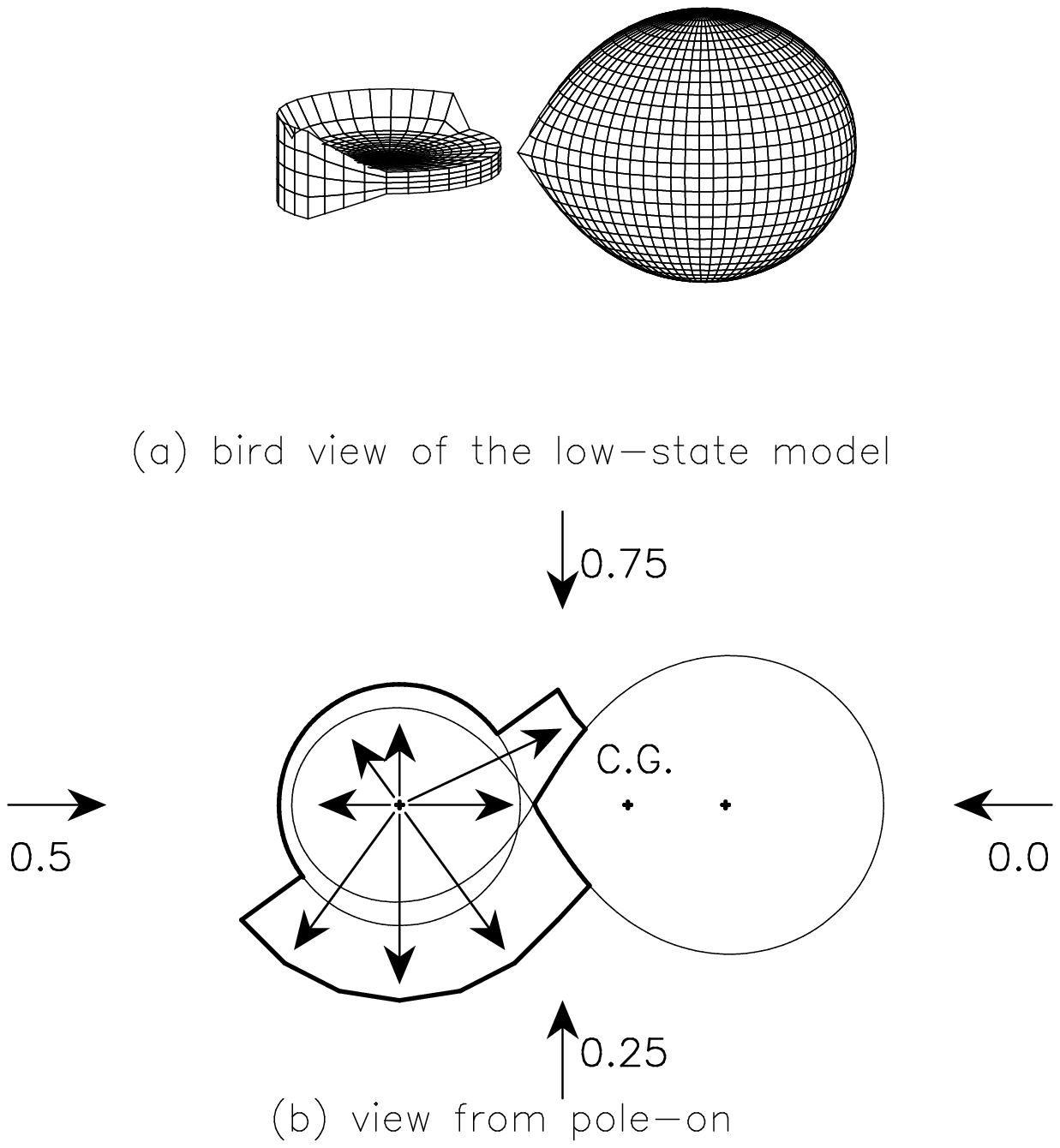}
\caption{
An optically thin, weak disk surface flow (wind) is driven 
by the central hot white dwarf 
when the white dwarf is very luminous \citep{fuk99}.
(a) The optically thick photo-surfaces (photospheres) are shown
in a bird view.
(b) The formation region of \ion{He}{2} 1640.476 emission line 
is illustrated (enclosed by a thick solid line).
Arrows in the line formation region indicate the velocities of 
the disk surface flow.  The disk surface flows can easily flow
out from phase 0.9 to 0.4 and are possibly 
accelerated up to 1500~km~s$^{-1}$ or more \citep{fuk99}.
On the other hand, it cannot be accelerated so much between phase
0.4 to 0.9 because the elevated edge of the disk hampers the 
acceleration.  The outer arrows represent the orbital phase 
(the primary eclipse is phase 0) viewed from the Earth.
\label{pole_on_view}}
\end{figure}

     \citet{woo00} explained the above features by the colliding
winds.  However, these features can be qualitatively explained
even by our WD model.  \citet{fuk99} showed that a disk surface
flow is driven up to $3000-4000$~km~s$^{-1}$ 
when the central WD is as luminous as the Eddington
luminosity.  The disk surface flow is line-driven, therefore, we see
an optically thin wind going outward along the disk surface.
When the disk edge
is asymmetric as illustrated in Figure \ref{pole_on_view}, 
the disk surface flow is also asymmetric because the elevated
edge of the disk suppresses effective acceleration.
(1) At phase zero, we can see both the elevated disk edge
and the extended emission region in the angle between phase 0.1
and phase 0.4 (see Fig. \ref{pole_on_view}).
The emission region of the elevated disk edge contributes 
to an excess at the red side of the central peak.
(2) As the binary phase increases, the blue side of the line
becomes more prominent.  During phase 0.1 to 0.4, the disk motion
is toward us and its velocity is increasing to $-300$~km~s$^{-1}$
at phase 0.25.  This orbital motion makes the sharp red edge of
the line blueward as shown in Figure 4 of \citet{woo00}. 
(3) At phase 0.4 the line is roughly symmetric but with an small
excess in the blue side.  This is very consistent with the 
observed one.  (4) From phase 0.5 to 0.8 the behavior of 
the line is opposite to that of phase 0.1 to 0.4 
as easily understood from Figure \ref{pole_on_view}.
Since the disk goes away from us during this phase,
the sharp blue edge of the line moves redward.  
This is also consistent with Figure 4 of \citet{woo00}. 
At phase 0.9 the line is almost symmetric again but with 
a very small excess in the red side and this is also
consistent with the observed line profiles.

\placefigure{disk_flow}

\subsection{The \ion{C}{4} doublet 1548 and 1551 in the low state} 
     The observational features of the \ion{C}{4} spectra
in the optical low states \citep{woo00} are summarized as follows:  
(1) At phase $0.5-0.7$ the blue
absorption feature belonging to the blue member of the
\ion{C}{4} doublet becomes largest with the P Cygni absorption
reaching $\sim -700$~km~s$^{-1}$.
(2) As the phase increases the extent decreases, reaching
only $\sim -250$~km~s$^{-1}$ just prior to eclipse.
(3) The extent of the absorption feature increases again
after eclipse to phase 0.5 but the absorption itself is rather weak.
\par
\begin{figure}
\plotone{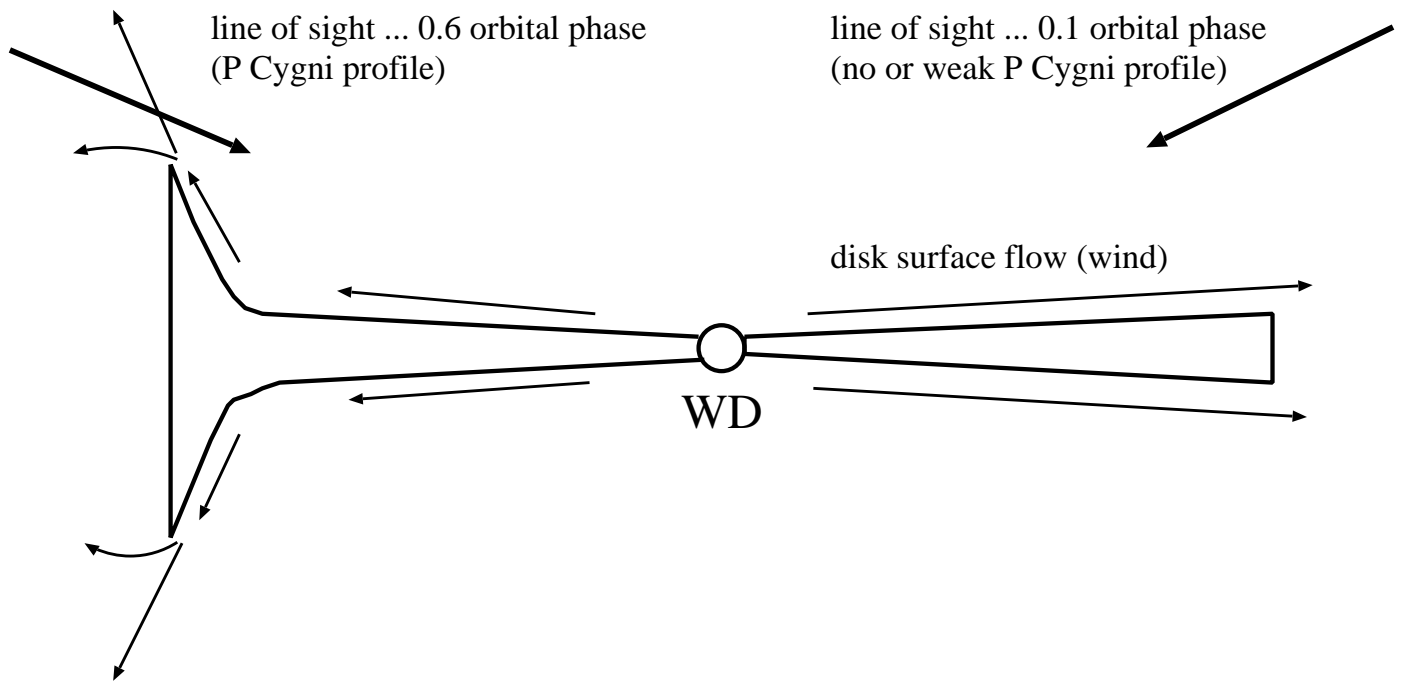}
\caption{
A cross sectional view of the disk surface flow (wind)
is shown.  No or weak self-absorption (P Cygni profile) of
the \ion{C}{4} doublet is observed
from the orbital phase of 0.1 while a rather strong self-absorption
is seen from the orbital phase of 0.6.  We assume that the line
formation region of the \ion{C}{4} doublet is more extended
compared with that of the \ion{He}{2} 1640 emission
in Figure \ref{pole_on_view}.
\label{disk_flow}}
\end{figure}

     To explain the above behavior we assume that the line 
formation region of \ion{C}{4} doublet is much more 
extended compared with that of \ion{He}{2} 1640.
An illustrated configuration of our model is shown in
Figure \ref{disk_flow}.  In the direction of 
phase 0.4 to 0.9, the disk surface flow goes up along
the elevated edge of the disk involving the disk matter
dragged.  Therefore, the disk surface flow
crosses the line of sight and absorbs the \ion{C}{4} lines.
On the other hand, in the direction of
phase 0.9 to 0.4 the disk surface flow virtually does not
cross the line of sight (see Fig. \ref{disk_flow}).
As a result, the absorption is very weak.
The disk surface flow has a large angle against the line of
sight, for example, $30-60\arcdeg$ or so.  Then, the extension
of the the P Cygni absorption feature is limited at
a smaller velocity like $\sim -700$~km~s$^{-1}$ even if its
terminal velocity is as large as $\sim 1000-1500$~km~s$^{-1}$.

\subsection{Other emission features in the low state} 
     Phase-resolved spectra of the 
\ion{He}{2} $\lambda 4686$ and H$\beta$ 
in the optical low states were obtained by \citet{dia99}.
The radial velocities of the emission lines were
measured by a double-Gaussian convolution method.  Their results 
are summarized as follows: \\
(1) The radial velocity amplitudes of
\ion{He}{2} $\lambda 4686$ and H$\beta$ agree with each other,
i.e., $K \sim 230$~km~s$^{-1}$, by Gaussian components half-separated
by $800$~km~s$^{-1}$ for \ion{He}{2} $\lambda 4686$ 
and by $1100$~km~s$^{-1}$ for H$\beta$. \\
(2) Positive-to-negative velocity crossing occurs at phase 0.93,
that is, the largest positive velocity is reached at phase 0.68
and the smallest negative velocity is attained at phase 0.18. \\
(3) The H$\beta$ line flux presents a sharp maximum during phase
0.0, which indicates that the Balmer line formation region is not
eclipsed, while the \ion{He}{2} $\lambda 4686$ line flux reaches
minimum at phase 0.0, indicating a partial eclipse of a highly
ionized region.
\par
     We attribute the \ion{He}{2} $\lambda 4686$ line formation
region to a relatively inner portion of the disk but
the H$\beta$ line formation region both to an outer portion of the disk
and the irradiated MS companion (see Fig. 
\ref{pole_on_view}).  The disk surface flow is accelerated by
the central WD and its outflow velocity gradually increases outward.
This tendency is roughly consistent with the Gaussian component
velocities of $800$~km~s$^{-1}$ (inner portion 
of the \ion{He}{2} $\lambda 4686$ line) 
and $1100$~km~s$^{-1}$ (outer portion of the H$\beta$).
\par
     The disk structure is asymmetric as shown in Figure 
\ref{pole_on_view}, so that we observe the largest positive
velocity at phase 0.65 (midpoint between phase 0.4 and 0.9)
and the lowest negative velocity at phase 0.15 if the disk 
surface flow is much faster than the orbital motion.
Adding the orbital motion of the WD to the disk outflow, 
the phase of the largest
positive velocity should be shifted a bit later
because the orbital motion reaches its
largest velocity at phase 0.75.  Therefore, phase 0.68 is
reasonable.  For the lowest negative velocity,
the phase should also be delayed from phase 0.15 because 
the orbital motion of the WD component reaches the lowest 
negative value at phase 0.25.  In this sense, phase 0.18
is also reasonable.
\par
     During phase 0.5 to 0.8, the blue shifted component 
of the disk flow is blocked by the elevated disk edge
as seen in Figure \ref{pole_on_view}, so that
an extended blue wing is probably cut in the spectra
as well as the self-absorption feature.
\citet{dia99} showed contour maps of phase-resolved 
line intensities both for the \ion{He}{2} $\lambda 4686$ 
and H$\beta$ lines. 
The above feature is clearly shown in Figures 4 and 5 of \citet{dia99}
both for the \ion{He}{2} $\lambda 4686$ and H$\beta$ lines.
\par
     Closely looking at the radial velocity change 
of the reddest contour in Figures 4 and 5 of \citet{dia99},
we easily find that the radial velocity reaches maximum of
$\sim 1500$~km~s$^{-1}$ at phase $\sim 0.7$ and minimum of
$\sim 900$~km~s$^{-1}$ at phase $\sim 0.1$, which suggests
an amplitude of $K_1 \sim 300$~km~s$^{-1}$ and
an outflow velocity of $\sim 1200$~km~s$^{-1}$.
These kinematic velocities are consistent with our WD wind model.
However, the bluest contour shows the minimum velocity
of $-1000$~km~s$^{-1}$ at phase 0.2 and the maximum velocity
of $-500$~km~s$^{-1}$ at phase 0.6 
for the \ion{He}{2} $\lambda 4686$ line.
This asymmetric feature is very similar to those of
\ion{He}{2} $\lambda 4686$ emission line in the optical high state
(see Fig. \ref{he2_line_high}), although the outflow
velocity is smaller in the low state.  It is therefore likely
that the blue edges of these emission lines are self-absorbed
even in the optical low state, unless it is significantly 
contaminated by \ion{C}{3}$+$\ion{N}{3} complex in the blue edge
over $-1000$~km~s$^{-1}$.
\par
     To summarize, our WD wind model possibly gives us natural
explanations of the observational features, i.e.,
\ion{He}{2} $\lambda 4686$ emission (together with H$\alpha$)
in the optical high states and
\ion{He}{2} $\lambda 1640$ emissions,
\ion{C}{4} doublet $\lambda 1548$ and 1551,
and \ion{He}{2} $\lambda 4686$ emission (together with H$\beta$)
in the optical low states, at least qualitatively.

\subsection{Weak soft X-ray flux} 
     \citet{dia95} pointed out 
that soft X-ray flux of V~Sge is too weak to be compatible
with the typical supersoft X-ray sources, for example, at least 
two or three orders of magnitude lower than that of CAL~87.
In this subsection, we try to resolve this problem.
Our numerical results are summarized as follows:
(1) the distance to V~Sge is 3~kpc, 
(2) the temperature of the WD is between $30-60$~eV
as seen in Figure \ref{vmag1250m35_long_time_fit_vsge},
and (3) the bolometric luminosity is $1.7 \times 10^{38}$~erg~s$^{-1}$.
If we adopt the color excess of $E(\bv)= 0.3$ and 
the corresponding hydrogen column density of 
$N_{\rm H} \sim 2 \times 10^{21}$~cm$^{-1}$ \citep{gor75},
the ROSAT/PSPC count rate is calculated 
to be larger than 2~counts~s$^{-1}$.
However, the observed rate is only 0.02~counts~s$^{-1}$ \citep{gre98},
at least, two orders of magnitude smaller than that of our WD model.
\par
     The central WD is always occulted by the elevated disk edge
in the CAL~87 model (model d) of \citet{sch97}.  
In this model,
a substantial part of soft X-rays can penetrate the elevated
edge of the disk, i.e., the spray is semi-transparent
at short wavelength \citep[see private communication of F. Meyer 
in][]{led03}.  Very recently, \citet*{sul03} formulated 
X-ray radiation reprocessing in the elevated edge of the disk.
In their results, some fraction of soft X-rays leak 
from the elevated edge that consists of many small
size cloud blobs and hot ($\sim 5 \times 10^{5}$~K) intercloud medium.
The mass transfer rate in CAL~87 is estimated
to be as small as $\sim 1 \times 10^{-7} M_\sun$~yr$^{-1}$
and two orders of magnitude
smaller than that for our V~Sge model.
It is very likely that the density of spray is much (e.g., 
two orders of magnitude) larger in V~Sge than in CAL~87. 
This much more opaque elevated disk edge results in
a very low flux of soft X-rays even if the
system configuration (orbital period and inclination) is almost
the same between these two systems.

\subsection{Galactic counterpart of RX~J0513.9$-$6951} 
      \citet{gre98} concluded that the model suggested for 
\objectname{RX J0513.9$-$6951} cannot explain 
the X-ray and optical variabilities
of \objectname{V Sge}.  Their reasons are as follows:
(1) both the transitions, from the faint to bright
and the bright to faint, are very rapid, so that they may occur 
in within one day (compared to the smooth
decline of several days in \objectname{RX J0513.9$-$6951}).
(2) The expected optical eclipse would become deeper in the brighter
state, opposite to what has been observed \citep{pat98}.
\par
     Their assumed model for \objectname{RX J0513.9$-$6951} is
based on the photospheric expansion/contraction model of the WD
envelope \citep[e.g.,][]{pak93, rei96, rei00}.  
Their conclusions suggest only that 
the photospheric expansion/contraction model does not apply to
\objectname{V Sge}.  As already shown in \citet{hac03ka, hac03kb},
the photospheric expansion/contraction model cannot explain
the X-ray variability even for \objectname{RX J0513.9$-$6951}
because the emergence/decay timescale of supersoft X-rays is also
as short as $\sim 1-2$ days \citep{rei00}.
\par
     On the other hand, our model can naturally explain the above two
observational features as follows:
(1) the irradiated disk is a main source 
of the optical light and its flaring-up edge is very variable
in a dynamical timescale, i.e., a day or so.  This also naturally
explain the optical variability from night to night 
\citep{loc99, mad97}.  The difference in the optical variabilities
between \objectname{V Sge} and \objectname{RX J0513.9$-$6951} may
come from the difference in the inclination angle, i.e.,
$i= 70-80\arcdeg$ for \objectname{V Sge} but $i= 20-30\arcdeg$ for
\objectname{RX J0513.9$-$6951} \citep{hut02, hac03kb}.
The irradiated area of the flaring-up disk viewed from 
the Earth varies much more largely in the binary system with
a higher inclination angle.  (2) As recently reported in
\citet{cow02}, the eclipse depth of \objectname{RX J0513.9$-$6951}
is deeper in the optical low state than in the optical high state.
This is the same as what has been observed in \objectname{V Sge}.
Our models can reproduce the orbital modulations both 
in the optical high and low states both for 
\objectname{V Sge} and \objectname{RX J0513.9$-$6951}.
Thus, we may conclude that \objectname{V Sge} is 
essentially the same type of objects as \objectname{RX J0513.9$-$6951},
and its Galactic counterpart.

\subsection{Relevance to Type Ia supernovae} 
     Thus, our optically thick wind model of mass-accreting white 
dwarfs can reproduce many observational features of V~Sge.
We suggest that V~Sge is the second example of 
accretion wind evolution, which is a key evolutionary process
to Type Ia supernovae in a recently developed 
evolutionary scenario \citep{hkn96, hkn99, hknu99, hac01kb}.
In the accretion wind evolution phase, the donor transfers 
mass to the white dwarf at a high rate of 
$\sim 1 \times 10^{-6} M_\sun$~yr$^{-1}$ or more.
In an old picture, the white dwarf expands to a giant size to form
a common envelope and results in merging.  
However, the white dwarf begins to blow a massive wind
in such a high mass accretion rate instead of simply expanding.
A formation of common envelope is avoided and the separation of
the binary is almost unchanged.  In the accretion wind evolution
phase, the white dwarf accretes matter and burns
hydrogen atop the white dwarf core.  Then, the white dwarf 
can grow in mass at the critical rate,
$\dot M_{\rm cr} \sim 1 \times 10^{-6} M_\sun$~yr$^{-1}$,
up to the Chandrasekhar mass limit and
explodes as a Type Ia supernova.  
With the first and second examples of
accretion wind evolution \objectname{RX J0513.9$-$6951} 
\citep{hac03ka, hac03kb} in the LMC 
and \objectname{V Sge} in our Galaxy,
we may conclude that accretion wind evolution is established in 
the evolutionary scenario to Type Ia supernovae.
In other words, accretion wind evolution commonly occurs in
the supersoft X-ray sources when the mass transfer rate exceeds
the critical rate of $\sim 1 \times 10^{-6} M_\sun$~yr$^{-1}$.
\par
     The life time of \objectname{V Sge} phenomena is 
as short as $\sim 10^{5}$~yr, because the mass transfer rate 
is as large as $\dot M_{\rm MS} \sim 10^{-5} M_\sun$~yr$^{-1}$.
\objectname{V Sge} belongs to the helium-rich supersoft 
X-ray source channel in the Type Ia supernova scenario of \citet{hknu99}.
This channel can produce about one Type Ia supernova per millennium,
indicating that we have a chance to observe about a hundred 
V Sagittae stars in our Galaxy.
\citet{ste98} have already listed four \objectname{V Sagittae} stars
in our Galaxy and discussed their similar properties.
Although the mass of the companion star to the WD
is not clearly identified yet, their orbital periods fall 
in the range of $0.2-0.5$ days, being very consistent with
the orbital periods that the new scenario predicted
\citep[see Fig. 3 of][]{hac01kb}.

\section{CONCLUSIONS}
     We have proposed a model of long-term optical variations
in the peculiar binary V~Sge based on an optically thick wind
model of mass-accreting white dwarfs.  
Our binary model, which is consisting
of a mass-accreting white dwarf (WD), an accretion disk around the
WD, and a lobe-filling main-sequence (MS) companion star,
reproduces orbital light curves both for the optical high and
low states.  Moreover, our WD wind model
naturally explains the transition between the optical high and 
low states. 
\par
     Here, we summarize our main results:\\
{\bf 1.} When the mass accretion rate to the white dwarf 
exceeds a critical rate, the white dwarf photosphere expands to
blow a fast and massive wind.  The disk surface is blown in the wind
and optically thick parts of the disk extend to a large size 
over the Roche lobe.  The largely extended disk, 
which is irradiated by a WD photosphere,
contributes much to the optical light.  This corresponds to
the optical high states. \\
{\bf 2.} The fast and massive wind hits the donor star and strips off
the surface layer.  The mass transfer rate from the
donor is attenuated.  As a result, the wind gradually weakens and stops.
Then the disk shrinks to a size of 
the Roche lobe.  This (no wind) phase corresponds 
to the optical low states because the disk irradiation area also 
becomes small. \\
{\bf 3.} In the optical bright state, we are able to reproduce
the orbital light curves of \objectname{V Sge}
with a large extension of the disk.
In the optical faint state, the orbital light curves 
are reproduced with a small size (similar size to the Roche lobe)
of the disk having an elevated edge as seen in the supersoft X-ray 
sources. \\
{\bf 4.} The massive wind easily shields supersoft X-rays.
Therefore, soft X-rays are not detected in the wind phase, i.e.,
in the optical high state.  This is consistent with 
the soft X-ray behavior found by \citet{gre98}. 
Furthermore, our wind model also explain the hard X-ray
component during the optical high state if they are shock-origin 
\citep{gre98}.  \\
{\bf 5.} The observed radio flux indicates a wind mass loss rate of
$\sim 10^{-5}M_\sun$~yr$^{-1}$ \citep{loc97}.  Our optically thick
wind model of mass-accreting white dwarfs
naturally explain such a high mass loss rate. \\
{\bf 6.} Various features of the emission lines both in the optical
high and low states are also naturally explained by our model. \\
{\bf 7.} Optically thick winds from mass-accreting WDs
play an essential role in a recently developed evolutionary
scenario of Type Ia supernovae \citep{hkn96, hkn99, hknu99}.
This process is called accretion wind evolution.
V~Sge is the second example of accretion wind evolution
with the first example of the LMC supersoft X-ray source
RX~J0513.9$-$6951 \citep{hac03ka, hac03kb}. 
With these two examples of accretion wind evolution,
we may conclude that optical high/low states with
wind being on/off are a common physical feature 
in the supersoft X-ray sources if the mass transfer rate
exceeds the critical rate of $\sim 1 \times 10^{-6} M_\sun$~yr$^{-1}$. \\ 
{\bf 8.} \objectname{V Sge} is a strong 
candidate for Type Ia supernova progenitors. \\



\acknowledgments
     We thank V. \v{S}imon for providing us machine readable
data of the AAVSO observation.
This research has been supported in part by the Grant-in-Aid for
Scientific Research (11640226) 
of the Japan Society for the Promotion of Science.

\begin{deluxetable}{llcrrrrll}
\tabletypesize{\scriptsize}
\tablecaption{Durations of high and low states
\label{high_low_states}}
\tablewidth{0pt}
\tablehead{
\colhead{$M_{\rm WD}$} & 
\colhead{$M_{\rm MS}$} & 
\colhead{$\dot M_{\rm MS}$} & 
\colhead{$t_{\rm vis}$} & 
\colhead{$c_1$} & 
\colhead{high} &
\colhead{low} &
\colhead{light} &
\colhead{comments} \\
\colhead{$(M_\sun)$} &
\colhead{$(M_\sun)$} &
\colhead{$(10^{-6} M_\sun$~yr$^{-1})$} &
\colhead{(day)} &
\colhead{} &
\colhead{(days)} &
\colhead{(days)} &
\colhead{curve} &
\colhead{}
} 
\startdata
0.8 & 2.5 & 5.0 & 13.9 & 10.0 & $\infty$ & \nodata &  & modulation\tablenotemark{a} \\
0.8 & 2.5 & 5.0 & 41.6 & 10.0 & $\infty$ & \nodata &  & modulation \\
1.0 & 3.0 & 5.0 & 41.6 & 10.0 & $\infty$ & \nodata &  & modulation \\
1.1 & 3.5 & 5.0 & 41.6 & 10.0 & 250 & 90 &  & \\
1.1 & 3.5 & 5.0 & 41.6 & 5.0 & $\infty$ & \nodata &  & modulation \\
1.1 & 3.5 & 5.0 & 69.4 & 5.0 & 310 & 30 &  & \\
1.1 & 3.5 & 5.0 & 69.4 & 1.5 & $\infty$ & \nodata &  & modulation \\
1.1 & 3.5 & 5.0 & 97.2 & 1.5 & $\infty$ & \nodata &  & modulation \\
1.1 & 3.5 & 5.0 & 111. & 1.5 & 390 & 10 &  &  \\
1.1 & 3.5 & 5.0 & 139. & 1.5 & 420 & 40 &  &  \\
1.1 & 3.5 & 5.0 & 194. & 1.5 & 500 & 100 &  &  \\
1.1 & 3.5 & 40. & 41.6 & 12.0 & 430 & 70 &  &  \\
1.1 & 3.5 & 40. & 41.6 & 13.0 & 430 & 110 &  &  \\
1.1 & 3.5 & 40. & 41.6 & 14.0 & 430 & 150 & Fig.\ref{vmag1100m35_long_time_fit_vsge}b &  \\
1.2 & 3.5 & 5.0 & 41.6 & 10.0 & 180 & 170 &  & \\
1.2 & 3.5 & 5.0 & 41.6 & 5.0 & 190 & 20 &  & \\
1.2 & 3.5 & 5.0 & 41.6 & 1.5 & $\infty$ & \nodata &  & modulation \\
1.2 & 3.5 & 5.0 & 69.4 & 1.5 & 230 & 10 &  &  \\
1.2 & 3.5 & 5.0 & 97.2 & 1.5 & 270 & 40 &  &  \\
1.2 & 3.5 & 5.0 & 139. & 1.5 & 320 & 90 &  &  \\
1.2 & 3.5 & 10. & 41.6 & 10.0 & 230 & 170 &  &  \\
1.2 & 3.5 & 10. & 41.6 & 7.0 & 230 & 60 &  &  \\
1.2 & 3.5 & 10. & 41.6 & 5.0 & $\infty$ & \nodata &  & modulation \\
1.2 & 3.5 & 20. & 41.6 & 10.0 & 270 & 150 &  &  \\
1.2 & 3.5 & 20. & 41.6 & 9.0 & 270 & 110 &  &  \\
1.2 & 3.5 & 20. & 41.6 & 7.0 & 270 & 40 &  &  \\
1.2 & 3.5 & 20. & 41.6 & 5.0 & $\infty$ & \nodata &  & modulation \\
1.2 & 3.5 & 30. & 41.6 & 9.0 & 280 & 100 &  &  \\
1.2 & 3.5 & 30. & 41.6 & 10.0 & 290 & 140 & Fig.\ref{vmag1200m35_long_time_fit_vsge}b &  \\
1.2 & 3.5 & 40. & 41.6 & 10.0 & 300 & 140 &  &  \\
1.2 & 3.5 & 40. & 41.6 & 9.0 & 300 & 100 &  &  \\
1.2 & 3.5 & 40. & 41.6 & 7.0 & 300 & 20 &  &  \\
1.2 & 3.5 & 40. & 41.6 & 5.0 & $\infty$ & \nodata &  & modulation \\
1.25 & 2.5 & 25. & 41.6 & 8.0 & 220 & 130 & Fig.\ref{vmag_ms_mass_long_time_fit_vsge}a & \\
1.25 & 3.0 & 25. & 41.6 & 8.0 & 220 & 130 & Fig.\ref{vmag_ms_mass_long_time_fit_vsge}b,d,e & \\
1.25 & 3.5 & 0.64 & 41.6 & 8.0 & 70 & 130 & Fig.\ref{vmag1250m35_long_time_fit_vsge_acc}a &  \\
1.25 & 3.5 & 1.0 & 41.6 & 1.5 & $\infty$ & \nodata &  & no modulation\tablenotemark{b} \\
1.25 & 3.5 & 1.0 & 41.6 & 8.0 & 80 & 120 & Fig.\ref{vmag1250m35_long_time_fit_vsge_acc}b &  \\
1.25 & 3.5 & 2.0 & 41.6 & 8.0 & 100 & 130 & Fig.\ref{vmag1250m35_long_time_fit_vsge_acc}c &  \\
1.25 & 3.5 & 5.0 & 41.6 & 8.0 & 150 & 140 & Fig.\ref{vmag1250m35_long_time_fit_vsge_acc}d &  \\
1.25 & 3.5 & 10. & 41.6 & 8.0 & 190 & 130 & Fig.\ref{vmag1250m35_long_time_fit_vsge_acc}e &  \\
1.25 & 3.5 & 20. & 41.6 & 8.0 & 220 & 120 & Fig.\ref{vmag1250m35_long_time_fit_vsge_acc}f &  \\
1.25 & 3.5 & 25. & 13.9 & 8.0 & $\infty$ & \nodata & Fig.\ref{vmag1250m35_long_time_fit_vsge_delay}a & modulation \\
1.25 & 3.5 & 25. & 27.8 & 8.0 & 200 & 30 & Fig.\ref{vmag1250m35_long_time_fit_vsge_delay}b & \\
1.25 & 3.5 & 25. & 41.6 & 1.5 & $\infty$ & \nodata &  & modulation \\
1.25 & 3.5 & 25. & 41.6 & 3.0 & $\infty$ & \nodata & Fig.\ref{vmag1250m35_long_time_fit_vsge_strip}a & modulation \\
1.25 & 3.5 & 25. & 41.6 & 5.0 & 220 & 10 & Fig.\ref{vmag1250m35_long_time_fit_vsge_strip}b & \\
1.25 & 3.5 & 25. & 41.6 & 8.0 & 220 & 130 &  Fig.\ref{vmag1250m35_long_time_fit_vsge}b & \\
1.25 & 3.5 & 25. & 41.6 & 8.0 & 230 & 120 &  & $\epsilon= 2 \times 10^{-7}$ \\
1.25 & 3.5 & 25. & 41.6 & 10.0 & 220 & 210 & Fig.\ref{vmag1250m35_long_time_fit_vsge_strip}d & \\
1.25 & 3.5 & 25. & 41.6 & 12.0 & 220 & 280 & Fig.\ref{vmag1250m35_long_time_fit_vsge_strip}e & \\
1.25 & 3.5 & 25. & 55.5 & 8.0 & 240 & 240 & Fig.\ref{vmag1250m35_long_time_fit_vsge_delay}d & \\
1.25 & 3.5 & 25. & 69.4 & 8.0 & 260 & 340 & Fig.\ref{vmag1250m35_long_time_fit_vsge_delay}e & \\
1.25 & 3.5 & 30. & 41.6 & 9.0 & 220 & 160 &  &  \\
1.25 & 3.5 & 40. & 41.6 & 8.0 & 230 & 130 & Fig.\ref{vmag1250m35_long_time_fit_vsge_acc}h &  \\
1.3 & 3.5 & 0.7 & 41.6 & 7.0 & 60 & 120 &  & \\
1.3 & 3.5 & 1.0 & 41.6 & 7.0 & 70 & 100 &  & \\
1.3 & 3.5 & 2.0 & 41.6 & 7.0 & 90 & 130 &  & \\
1.3 & 3.5 & 5.0 & 41.6 & 10.0 & 130 & 220 &  & \\
1.3 & 3.5 & 5.0 & 41.6 & 7.0 & 130 & 140 &  & \\
1.3 & 3.5 & 5.0 & 41.6 & 5.0 & 130 & 80 &  & \\
1.3 & 3.5 & 5.0 & 41.6 & 1.5 & 130 & 20 &  & \\
1.3 & 3.5 & 5.0 & 69.4 & 1.5 & 160 & 40 &  & \\
1.3 & 3.5 & 5.0 & 125. & 1.5 & 220 & 100 &  & \\
1.3 & 3.5 & 10. & 41.6 & 10.0 & 150 & 240 &  & \\
1.3 & 3.5 & 10. & 41.6 & 7.0 & 150 & 140 &  & \\
1.3 & 3.5 & 10. & 41.6 & 5.0 & 150 & 70 &  & \\
1.3 & 3.5 & 10. & 41.6 & 1.5 & 150 & 10 &  & \\
1.3 & 3.5 & 20. & 41.6 & 10.0 & 170 & 250 &  & \\
1.3 & 3.5 & 20. & 41.6 & 7.0 & 170 & 130 & Fig.\ref{vmag1300m35_long_time_fit_vsge}b & \\
1.3 & 3.5 & 20. & 41.6 & 5.0 & 170 & 60 &  & \\
1.3 & 3.5 & 20. & 41.6 & 1.5 & 170 & 10 &  & \\
1.3 & 3.5 & 40. & 41.6 & 10.0 & 180 & 260 &  & \\
1.3 & 3.5 & 40. & 41.6 & 7.0 & 180 & 140 &  & \\
1.3 & 3.5 & 40. & 41.6 & 5.0 & 180 & 60 &  & \\
1.3 & 3.5 & 40. & 41.6 & 1.5 & 180 & 10 &  & \\
\enddata
\tablenotetext{a}{mass accretion rate $\dot M_{\rm acc}$ 
is modulated by winds}
\tablenotetext{b}{mass accretion rate is slightly attenuated but
not modulated largely by winds, so that it is almost constant in time}
\end{deluxetable}






\clearpage

\clearpage

\clearpage

\clearpage

\clearpage

\clearpage

\clearpage

\clearpage

\clearpage

\clearpage

\clearpage

\clearpage

\clearpage

\clearpage

\clearpage



\begin{thebibliography}{}
\bibitem[Alcock et al. (1996)]{alc96}
Alcock, C. et al. 1996, \mnras, 280, L49


\bibitem[Bressan et al. (1993)]{bre93}
Bressan, A., Fagotto, F., Bertelli, G., \& Chiosi, C. 1993, \aaps, 100, 647

\bibitem[Cowley et al. (2002)]{cow02}
Cowley, A.P., Schmidtke, P.C., Crampton, D., \& Hutchings, J.B. 2002, 
\aj, 124, 2233

\bibitem[Cowley et al. (1993)]{cow93}
Cowley, A. P., Schmidtke, P. C., Hutchings, J. B., Crampton, D., \& 
McGrath, T. K. 1993, \apj, 418, L63

\bibitem[Crampton et al. (1996)]{cra96}
Crampton, D., Hutchings, J. B., Cowley, A. P., Schmidtke, P. C.,
McGrath, T. K., O'Donoghue, D., \& Harrop-Allin, M. K. 1996, \apj, 
456, 320

\bibitem[Diaz (1999)]{dia99}
Diaz, M. P. 1999, \pasp, 111, 76

\bibitem[Diaz \& Steiner (1995)]{dia95}
Diaz, M. P., \& Steiner, J. E. 1995, \aj, 110, 1816 


\bibitem[Fukue \& Hachiya (1999)]{fuk99}
Fukue, J., \& Hachiya, M. 1999, \pasj, 51, 185

\bibitem[G\"ansicke et al. (1998)]{gan98}
G\"ansicke, B. T., van Teeseling, A., Beuermann, K., \& de Martino, D.
1998, \aap, 333, 163

\bibitem[Gies et al. (1998)Gies, Shafter, \& Wiggs]{gie98}
Gies, D. R., Shafter, A. W., \& Wiggs, M. S. 1998, \aj, 115, 2566



\bibitem[Greiner \& van Teeseling (1998)]{gre98}
Greiner, J., \& van Teeseling, A. 1998, \aap, 339, L21

\bibitem[Gorenstein (1975)]{gor75}
Gorenstein, P. 1975, \apj, 198, 95

\bibitem[Hachisu \& Kato (2001a)]{hac01ka}
Hachisu, I., \& Kato, M. 2001a, \apjl, 553, L161

\bibitem[Hachisu \& Kato (2001b)]{hac01kb}
Hachisu, I., \& Kato, M. 2001b, \apj, 558, 323

\bibitem[Hachisu \& Kato (2003a)]{hac03ka}
Hachisu, I., \& Kato, M. 2003a, \apj, 588, 1003

\bibitem[Hachisu \& Kato (2003b)]{hac03kb}
Hachisu, I., \& Kato, M. 2003b, \apj, 590, 445



\bibitem[Hachisu et al. (1996)Hachisu, Kato, \& Nomoto]{hkn96}
Hachisu, I., Kato, M., \& Nomoto, K. 1996, \apj, 470, L97 

\bibitem[Hachisu et al. (1999a)Hachisu, Kato, \& Nomoto]{hkn99}
Hachisu, I., Kato, M., \& Nomoto, K. 1999a, \apj, 522, 487 

\bibitem[Hachisu et al. (1999b)]{hknu99}
Hachisu, I., Kato, M., Nomoto, K., \& Umeda, H. 1999b, \apj, 519, 314 

\bibitem[Hachisu et al. (2003)Hachisu, Kato, \& Schaefer]{hac03a}
Hachisu, I., Kato, M., \& Schaefer, B. E. 2003, \apj, 584, 1008
%



\bibitem[Herbig et al. (1965)]{her65}
Herbig, G. H., Preston, G. W., Smak, J., \& Paczynski, B. 1965, \apj,
141, 617

\bibitem[Hoard et al. (1996)Hoard, Wallerstein, \& Willson]{hoa96}
Hoard, D. W., Wallerstein, G., \& Willson, L. A. 1996, \pasp, 108, 81

\bibitem[Hutchings et al. (2002)]{hut02}
Hutchings, J. B., Winter, K., Cowley, A. P., Schmidtke, P. C.,
\& Crampton, D. 2002, \aj, 124, 2833


\bibitem[Kato (1983)]{kat83}
Kato, M. 1983, \pasj, 35, 507


\bibitem[Kato \& Hachisu (1994)]{kat94h}
Kato, M., \& Hachisu, I., 1994, \apj, 437, 802



\bibitem[Koch et al. (1986)]{koc86}
Koch, R. H., Corcoran, M. F., Holenstein, B. D., McCluskey, G. E. Jr.
1986, \apj, 306, 618

\bibitem[Langer et al. (2000)]{lan00}
Langer, N., Deutschmann, A., Wellstein, S., \& H\"oflich, P. 2000,
\aap, 362, 1046

\bibitem[Lederle \& Kimeswenger (2003)]{led03}
Lederle, C., \& Kimeswenger, S. 2003, \aap, 397, 951

\bibitem[Li \& van den Heuvel (1997)]{lih97}
Li, X.-D., \& van den Heuvel, E. P. J. 1997, \aap, 322, L9

\bibitem[Lockley (1998)]{loc98dt}
Lockley, J. J. 1998, Ph. D. thesis, Univ. Keele

\bibitem[Lockley et al. (1997)Lockley, Eyres, \& Wood]{loc97}
Lockley, J. J., Eyres, S. P. S., \& Wood, Janet H. 1997, \mnras, 287, L14

\bibitem[Lockley \& Wood (1998)]{loc98}
Lockley, J. J., \& Wood, J. H. 1998, in ASP Conf. Ser. 137, 13th North
American Work Shop on Cataclysmic Variables, ed. S. Howell, E.
Kuulkers, \& C. Woodward (San Francisco: ASP), 461

\bibitem[Lockley et al. (1999)]{loc99}
Lockley, J. J., Wood, J. H., Eyres, S. P. S., Naylor, T., \&
 Shugarov, S. 1999, \mnras, 310, 963



\bibitem[Mader \& Shafter (1997)]{mad97}
Mader, J., \& Shafter, A. 1997, \pasp, 109, 1351





\bibitem[Pakull et al. (1993)]{pak93}
Pakull, M. W., Moch, C., Bianchi, L., Thomas, H.-C., Guibert, J., 
Beaulieu, J. P., Grison, P., \& Schaeidt, S. 1993, \aap, 278, L39

\bibitem[Patterson et al. (1998)]{pat98}
Patterson, J. et al. 1998, \pasp, 110, 380



\bibitem[Reinsch et al. (1996)]{rei96}
Reinsch, K., van Teeseling, A., Beuermann, K., \& Abbott, T. M. C. 1996,
\aap, 309, L11

\bibitem[Reinsch et al. (2000)]{rei00}
Reinsch, K., van Teeseling, A., King, A. R., \& Beuermann, K. 2000,
\aap, 354, L37


\bibitem[Ritter et al. (2000)Ritter, Zhang, \& Kolb]{rit00}
Ritter, H., Zhang, Z.-Y., \& Kolb, U. 2000, \aap, 360, 959

\bibitem[Robertson et al. (1997)Robertson, Honeycutt, \& Pier]{rob97}
Robertson, J. W., Honeycutt, R. K., \& Pier, J. R. 1997, \aj, 113, 787




\bibitem[Schaeidt et al. (1993)Schaeidt, Hasinger, \& Truemper]{sch93}
Schaeidt, S., Hasinger, G., \& Truemper, J. 1993, \aap, 270, L9

\bibitem[Schandl et al. (1997)Schandl, Meyer-Hofmeister, \& Meyer]{sch97}
Schandl, S., Meyer-Hofmeister, E., \& Meyer, F. 1997, \aap, 318, 73


\bibitem[Shakura \& Sunyaev (1973)]{sha73}
Shakura, N. I., \& Sunyaev, R. A. 1973, \aap, 24, 337

\bibitem[\v{S}imon (1996a)]{sim96a}
\v{S}imon, V. 1996a, \aaps, 118, 421

\bibitem[\v{S}imon (1996b)]{sim96b}
\v{S}imon, V. 1996b, \aap, 309, 775

\bibitem[\v{S}imon et al. (2002)]{sim02}
\v{S}imon, V., Hric, L., Petr\'ic, K., Shugarov, S.,
 Niarchos, P., \& Marsakova, V. I. 2002, \aap, 393, 921

\bibitem[\v{S}imon \& Mattei (1999)]{sim99}
\v{S}imon, V., \& Mattei, J. A. 1999, \aaps, 139, 75

\bibitem[\v{S}imon et al. (2001)]{sim01}
\v{S}imon, V., Shugarov, S., Marsakova, V. I. 2001, \aap, 366, 100

\bibitem[Smak (1995)]{sma95}
Smak, J. I. 1995, Acta Astr., 45, 361

\bibitem[Smak et al. (2001)Smak, Belczynski, \& Zola]{sma01}
Smak, J. I., Belczynski, K., \& Zola, S. 2001, Acta Astr., 51, 117

\bibitem[Southwell et al. (1996)]{sou96}
Southwell, K. A., Livio, M., Charles, P. A., O'Donoghue, D., 
\& Sutherland, W. J. 1996, \apj, 470, 1065


\bibitem[Steiner \& Diaz (1998)]{ste98}
Steiner, J. E., \& Diaz, M. P. 1998, \pasp, 110, 276

\bibitem[Suleimanov et al. (2003)Suleimanov, Meyer, \& Meyer-Hofmeister]{sul03}
Suleimanov, V., Meyer, F., \& Meyer-Hofmeister, E. 2003, \aap, 
401, 1009




\bibitem[van den Heuvel et al. (1992)]{heu92}
van den Heuvel, E. P. J., Bhattacharya, D., Nomoto, K., \& Rappaport, 
S. 1992, \aap, 262, 97



\bibitem[Williams et al. (1986)]{wil86}
Williams, G. A., King, A. R., Uomoto, A. K., \& Hiltner, W. A. 1986,
\mnras, 219, 809

\bibitem[Wood \& Lockley (1997)]{woo97}
Wood, J. H., \& Lockley, J. J. 1997, in IAU Colloq. 163, Accretion
Phenomena and Related Outflows, ed. D. T. Wickramasinghe, G. V. 
Bicknell, \& L. Ferrario (ASP Conf. Ser. 121) (San Francisco: ASP),
457

\bibitem[Wood \& Lockley (2000)]{woo00}
Wood, J. H., \& Lockley, J. J. 2000, \mnras, 313, 789
%
\end{thebibliography}
\end{document}